\documentclass{elsart}

\usepackage{amsfonts}
\usepackage{amssymb}
\usepackage{graphicx}
\usepackage{psfrag}

\newcommand{\M}{\mathcal{M}}

\newcommand{\N}{\mathcal{N}}

\newcommand{\bn}{\mathbb{N}}
\newcommand{\br}{\mathbb{R}}

\newcommand{\p}{\varphi}
\newcommand{\s}{\psi}

\newcommand{\rsfp}{{\ell^*_{(\M,\varphi)}}}
\newcommand{\rsfs}{{\ell^*_{(\M,\psi)}}}
\newcommand{\rsfn}{{\ell^*_{(\N,\psi)}}}

\begin{document}
\begin{frontmatter}

\title{Natural pseudo-distance and optimal matching between reduced size functions}

\author[arces]{Michele d'Amico}
\author[arces,dip]{Patrizio Frosini}
\author[arces,dismi]{Claudia Landi\corauthref{cor}}
\corauth[cor]{Corresponding author: {\it E-mail Address:} {\tt clandi@unimore.it} (Claudia Landi)}
\address[arces]{ARCES, Universit\`a di Bologna, via Toffano 2/2, I-40135 Bologna, Italia}
\address[dip]{Dipartimento di Matematica, Universit\`a di Bologna, P.zza di Porta San Donato 5, I-40127 Bologna,
        Italia}
\address[dismi]{DISMI, Universit\`a di Modena e Reggio Emilia, via Amendola 2 - Pad. Morselli, I-42100 Reggio Emilia, Italia}

\begin{abstract}
This paper studies the properties of a new lower bound for the natural pseudo-distance. The natural pseudo-distance is 
 a dissimilarity measure between shapes,  
where a shape is viewed as a topological space endowed with a
real-valued continuous function. Measuring dissimilarity amounts to minimizing the change in the functions due to the
application of homeomorphisms between topological spaces, with
respect to the $L_\infty$-norm. In order to obtain the lower bound, a suitable metric between size functions, called matching distance, is introduced. It compares size functions  by solving an optimal
matching problem between countable point sets. The matching  distance is shown to be resistant to perturbations, implying that it is always smaller than the natural pseudo-distance. We also prove that the lower bound so obtained is sharp and cannot be improved by any other distance between size functions.
\end{abstract}


\end{frontmatter}

\section{Introduction}

Shape comparison is a fundamental problem in shape recognition, shape
classification and shape retrieval (cf., e.g.,  \cite{PR}), finding its applications mainly in Computer Vision and Computer Graphics.  The shape
comparison problem is  often  dealt with by defining a suitable
distance providing a measure of dissimilarity between shapes (see,
e.g., \cite{Veltkamp}
 for a review of the literature).

Over the last fifteen years, Size Theory has been developed as a
geometrical-topological theory for comparing shapes, each shape
viewed as a topological space $\M$, endowed with a real-valued
continuous function $\varphi$ \cite{FroLa99}.
The
pair $(\M,\varphi)$ is called a {\em size pair}, while $\varphi$
is said to be a {\em measuring function}. The
role of the  function $\p$ is to take into account only the shape
properties of the object described by $\M$ that are relevant to
the shape comparison problem at hand, while disregarding the
irrelevant ones, as well as to impose the desired invariance
properties.

A measure of the dissimilarity between two size pairs is  given by the {\em natural pseudo-distance}.
The main idea in the definition of natural pseudo-distance between
size pairs  is to minimize the change in the measuring functions
due to the application of homeomorphisms between topological
spaces, with respect to the $L_\infty$-norm: The natural pseudo-distance between $(\M,\varphi)$ and $(\N,\psi)$ with $\M$ and $\N$ homeomorphic is the number 
$$\inf_{h}\max _{P\in
\M}|\p(P)-\s(h(P))|,$$
where $h$ varies in the set $H(\M,\N)$ of all the homeomorphisms between $\M$ and $\N$.
In other words, the variation of the shapes is modeled by the infinite-dimensional group of homeomorphisms and  the cost of warping an object's shape into another is measured by the change of the measuring functions. An important feature of the natural pseudo-distance is that it does not require the choice of parametrizations for the spaces under study nor the choice of origins of coordinate, which in image applications would be arbitrarily driven. 

The main aim of this paper is to provide a new method to estimate the natural pseudo-distance, motivated by the intrinsic difficulty of a direct computation. Since the  natural pseudo-distance is defined by a minimization process, it would be natural to look  for  the optimal transformation
that takes one shape into the other, as usual in energy minimization methods. In our case, however,  the existence of an optimal homeomorphism attaining the natural pseudo-distance is not guaranteed. 

Earlier results about the natural pseudo-distance can be divided in two classes. One class provides constraints on the possible values taken by the natural pseudo-distance between two size pairs. For example,  if the considered topological spaces $\M$ and $\N$ are smooth closed manifolds and the measuring functions are also smooth, then  the natural pseudo-distance is an integer sub-multiple of the Euclidean distance between two suitable critical  values of the measuring functions  \cite{donatini2004}. In particular,  this integer can only be either $1$ or $2$ in the case of curves, while it can be either $1$, $2$ or $3$ in the case of surfaces  \cite{donatini2007}. The other class of results furnishes lower bounds for the natural pseudo-distance \cite{Donatini}, \cite{FroMu99}. 
In particular
it is possible to estimate the natural pseudo-distance by using the concept of {\em size function} \cite{Donatini}. 
Indeed, size functions can
reduce the comparison of shapes to  the comparison of certain countable subsets of the real plane (cf. \cite{york},
\cite{sdiego} and \cite{landi2001}). This reduction allows us to
study the space of all homeomorphisms between the considered
topological spaces, without actually computing them.
The research on size functions has led to a formal setting, which has
turned out to be useful, not only from a theoretical
point of view, but also on the applicative side (see, e.g., \cite{EVAM}, \cite{CeFeGi06}, 
\cite{d'Amico2000}, \cite{Dibos},
 \cite{morftransf}, \cite{fros-pit99}, \cite{Ziou}). 
 
This paper investigates into the problem of obtaining lower bounds for the natural pseudo-distance using size functions. 

To this aim,   
we first introduce the
concept of {\em reduced size function}. Reduced size functions are a slightly modified version of size functions based on the connectedness relation instead of path-connectedness. This new definition
is introduced in order to obtain both
theoretical and computational advantages (see Rem.~\ref{rightcont} and Rem.~\ref{adv}).
However, the main properties of size functions are maintained. In particular, reduced size functions can be represented by  sets of points of the (extended) real plane, called {\em cornerpoints}.

Then we need a preliminary result about reduced size functions (Th.~\ref{main}).  It states that a suitable distance between reduced size functions exists,
 which is continuous with respect to  the measuring functions (in the sense of the $L_\infty$-topology). We call this distance
{\em matching distance}, since the underlying idea is to measure the cost of matching the two sets of cornerpoints
describing the reduced size functions. The matching distance reduces to the {\em bottleneck distance} used in \cite{CoEdHa07} for comparing Persistent Homology Groups when the measuring functions are taken in the subset of {\em tame functions}.   We underline that the continuity of the matching distance implies a property of perturbation robustness  for size functions allowing them to be used
in real applications.

Having proven this, we are ready to obtain our  main results.   Indeed, the stability of the matching distance allows us
to prove a sharp lower bound for the change of measuring functions
under the action of homeomorphisms between topological spaces, i.e. for the natural pseudo-distance (Th.~\ref{nsd}).  
Furthermore, we prove that the lower bound obtained using the matching distance not only improves the previous known lower bound stated in \cite{Donatini}, but is the best possible lower bound for the natural pseudo-distance obtainable using size functions. The proof of these facts is based on  Lemma~\ref{cubi}. This lemma is a crucial result stating that it is always possible to construct two suitable measuring functions on a topological $2$-sphere
with given reduced size functions and a pseudo-distance
equaling their matching distance. On the basis of this lemma, in
Th.~\ref{ermejo} and Th. \ref{confronto} we prove that the matching distance we
are considering is, in two different ways, the best metric to compare reduced size functions.

This paper is organized as follows. In Section~2 we
introduce the concept of reduced size function and its main
properties. In Section~3 the definition of matching distance
between reduced size functions is given. In Section~4 the
stability theorem  is proved, together with some other useful
results. The connection with natural pseudo-distances between size
pairs is shown in Section~5, together with the proof of the
existence of an optimal matching between reduced size functions.
Section~6 contains the proof that it is always possible to
construct two size pairs with pre-assigned  reduced size functions
and a pseudo-distance equaling their matching distance. This
result  is used in Section~7 to conclude that the matching
distance furnishes the finest lower bound for the  natural
pseudo-distance between size pairs among the lower bounds
obtainable through reduced size functions. In Section~8 our results are briefly discussed.

\section{Reduced size functions}
In this section we   introduce reduced size functions, that is,  a
 notion derived from  size functions (\cite{FroLa99}) allowing for
 a simplified treatment of the theory. The definition of reduced size function differs from that of size function in that it is based on the relation of connectedness rather than on path-connectedness.  The motivation for this change,  as explained in Remark \ref{rightcont}, has to do with the right-continuity of size functions.

In what follows, $\M$ denotes a non-empty compact  connected and locally  connected Hausdorff space, representing the object whose shape is under study.

The assumption on the connectedness of $\M$ can easily be weakened
to any finite number of connected components without much affecting
the following results. More serious problems would derive from
considering an infinite number of connected components.

We shall call any pair $(\M,\varphi)$, where
$\varphi :\M\rightarrow \br$ is a continuous function,
a {\it size pair}. The function $\varphi$ is said to be a {\it measuring
function}. The
role of the  function $\p$ is to take into account only the shape
properties of the object described by $\M$ that are relevant to
the shape comparison problem at hand, while disregarding the
irrelevant ones, as well as to impose the desired invariance
properties.

Assume a size pair $({\M},\varphi)$ is given. For every
$x\in\mathbb{R}$, let ${\M}\langle\varphi\leq x\rangle$ denote the
lower level set $\{ P\in{\M} :\varphi (P)\leq x\}$. 

\begin{defn}
For every real number $y$, we shall say that two points $P,Q\in
{\M}$ are $\langle\varphi\leq y\rangle${\em -connected} if and
only if a connected subset $C$ of ${\M}\langle\varphi\leq
y\rangle$  exists, containing both $P$ and $Q$.
\end{defn}

In the following, $\Delta$ denotes the diagonal $\{(x,y)\in\br^2:x=y\}$; $\Delta^+$
denotes the open half-plane $\{(x,y)\in\br^2:x<y\}$  above the
diagonal; ${\bar \Delta^+}$
 denotes the closed
half-plane $\{(x,y)\in\br^2:x\le y\}$ above the diagonal.

\begin{defn}{\rm (Reduced size function)}
We shall call {\em reduced  size function} associated with the size
pair $(\M,\p)$ the  function $\rsfp:\Delta^+\rightarrow \bn$,
defined by setting $\ell^* _{({\M} ,\varphi )}(x,y)$ equal to the
number of equivalence classes into which the set
${\M}\langle\varphi\leq x\rangle$ is divided by the relation of
$\langle\varphi\leq y\rangle$-connectedness.
\end{defn}

In other words,    $\ell^* _{({\M} ,\varphi
)}(x,y)$ counts the number of connected components in
${\M}\langle\varphi\leq y\rangle$ that contain at least one point
of ${\M}\langle\varphi\leq x\rangle$. The finiteness of this
number is an easily obtainable consequence of the  compactness and
local-connectedness of $\M$.

\par
\begin{figure}
\psfrag{P}{$P$}
\psfrag{a}{$a$}
\psfrag{b}{$b$}
\psfrag{c}{$c$}
\psfrag{e}{$e$}
\psfrag{x}{$x$}
\psfrag{y}{$y$}
\psfrag{0}{$0$}
\psfrag{1}{$1$}
\psfrag{2}{$2$}
\psfrag{3}{$3$}
\centerline{
 \includegraphics[width=4in]{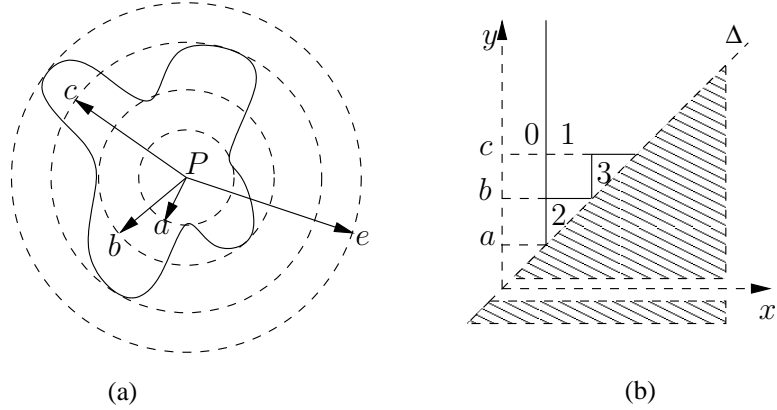}
} \caption{(b) The reduced size function of the  size pair  $(\M,\p)$,
where  $\M$ is the curve represented by a continuous line in (a),
and $\p$ is the function ``distance from the point
$P$''.}\label{Fig:sf}
\end{figure}
\par

An example of reduced size  function is illustrated in
Fig.~\ref{Fig:sf}. In this  example we consider the size pair
$(\M,\p)$, where $\M$ is the curve represented by a continuous
line in Fig.~\ref{Fig:sf}~(a), and $\p$ is the function ``distance
from the point $P$''.  The reduced size function associated with
$(\M,\p)$ is shown in Fig.~\ref{Fig:sf}~(b). Here, the domain
of the reduced size function is divided by solid lines,
representing the discontinuity points of the reduced size
function. These discontinuity points divide $\Delta^+$ into
regions in which the reduced size function is constant. The value
displayed in each region is the value taken by  the reduced size
function in that region.

For instance, for $a\le x<b$, the set $\{P\in \M:\p(P)\le x\}$ has two
connected components which  are contained in different
connected components of $\{P\in \M:\p(P)\le y\}$ when $x< y<b$.
Therefore, $\ell^*_{\left({\M},\varphi\right)}(x,y)=2$ for $a\le x<b$
and $x< y<b$.
When $a\le x<b$ and  $y\ge b$, all the connected components of $\{P\in \M:\p(P)\le x\}$ are contained in the same
connected component of $\{P\in \M:\p(P)\le y\}$.
Therefore, $\ell^*_{\left({\M},\varphi\right)}(x,y)=1$ for $a\le x<b$
and $y\ge b$.
When $b\le x<c$ and $y\ge c$, all of the three connected
components of $\{P\in \M:\p(P)\le x\}$ belong to the same
connected component of $\{P\in \M:\p(P)\le y\}$, implying that in
this case $\ell^*_{\left({\M},\varphi\right)}(x,y)=1$.

As for the values taken  on the discontinuity lines, they are easily obtained by observing
that reduced size functions are right-continuous, both in the variable $x$ and in the variable $y$.

\begin{rem}\label{rightcont}{\em 
 The property of right-continuity in the variable $x$ can  easily be checked and holds for classical size functions as well. The analogous property for the variable $y$ is not immediate, and  in general does not hold for classical size functions,  if not under stronger assumptions, such as, for instance,  that  $\M$ is a smooth manifold and the measuring function is  Morse (cf. Cor.~2.1 in \cite{frosini96}). Indeed, the relation of $\langle \p\le y\rangle$-homotopy, used to define classical size functions, does not pass to the limit. On the contrary, the relation of
$\langle\varphi\leq y\rangle$-connectedness does, that is to say, if, for every  $\epsilon>0$ it holds that $P$ and $Q$ are $\langle \varphi\leq
y+\epsilon\rangle$-connected, then they are $\langle\varphi\leq y\rangle$-connected. To see this, observe that connected components are closed sets, and the intersection $\bigcap_i K_i$ of a family of  compact, connected  Hausdorff subspaces $K_i$ of  a topological space, with the property that $K_{i+1}\subseteq K_i$ for every $i$, is connected (cf. Th.~28.2 in \cite{willard} p.~203).}
\end{rem}

Most properties of classical size functions continue to hold for
reduced size functions. For the aims of this paper, it is important that  
for reduced size  functions it is possible to define an analog of  classical size functions' cornerpoints and cornerlines, here respectively called proper cornerpoints and cornerpoints at infinity. The main reference here is \cite{landi2001}.

\begin{defn}{\rm (Proper cornerpoint)}
\label{cornerpt} For every point $p=(x,y)\in\Delta^+$, let us define the  number $\mu (p)$ as the minimum over all the positive  real numbers $\epsilon$,  with $x+\epsilon
<y-\epsilon$, of
$$\ell^* _{({\M},\varphi )}(x+\epsilon ,y-\epsilon )-\ell^* _{({\M},\varphi )}(x-\epsilon ,y-\epsilon )-\ell^* _{({\M},\varphi)} (x+\epsilon ,y+\epsilon )+\ell^* _{({\M},\varphi )}(x-\epsilon ,y+\epsilon ).$$
The finite number  $\mu (p)$ will be
called {\em multiplicity of} $p$ for $\ell^*  _{({\M},\varphi )}$.
Moreover,  we shall call {\em proper cornerpoint} for $\ell^*  _{({\M},\varphi )}$
any point $p\in\Delta^+$ such that  the number $\mu (p)$ is strictly
positive.
\end{defn}

\begin{defn}{\rm (Cornerpoint at infinity)}
\label{cornerptinfty}
For every vertical line $r$, with equation $x=k$, let us identify $r$ with the pair $(k,\infty)$, and define the number
$\mu(r)$ as the minimum, over all the positive real numbers $\epsilon$ with $k+\epsilon<1/\epsilon$, of
$$\ell ^*_{({\M},\varphi )}(k+\epsilon,1/\epsilon)-
\ell ^*_{({\M},\varphi )}(k-\epsilon,1/\epsilon).$$
When this finite number, called {\em multiplicity of} $r$ for $\ell^*
_{({\M},\varphi )}$, is strictly positive, we call the line $r$ a {\em
cornerpoint at infinity} for the reduced size function.
\end{defn}

\begin{rem}{\em 
The multiplicity of points and of vertical lines is always non negative. This follows from an analog of Lemma~1 in \cite{landi2001}, based on counting the equivalence classes in the  set
$$
\left\{ P\in {\M}\langle\varphi\leq x_2\rangle : \not\!\exists
Q\in {M}\langle\varphi\leq x_1 \rangle \ s.t.\ P
\cong_{\varphi \leq y_1}Q\right\}
$$
quotiented by the relation of $\langle\p\le y_1\rangle$-connectedness, in order to obtain the number
$$\rsfp(x_2,y_1)-\rsfp(x_1,y_1)$$
when $x_1\le x_2<y_1$.}
\end{rem}

\begin{rem}{\em 
Under our assumptions on $\M$, i.e.  its connectedness, $\mu(r)$ can only take  the values
$0$ and $1$, but the definition can easily  be extended to
spaces with any finite number of connected components, so that $\mu(r)$ can equal any natural number. Moreover, the connectedness assumption also implies  that there is exactly one cornerpoint at infinity.}
\end{rem}

\begin{figure}
\psfrag{x}{$x$}
\psfrag{y}{$y$}
\psfrag{m}{$m$}
\psfrag{p}{$p$}
\psfrag{q}{$q$}
\psfrag{s}{$s$}
\psfrag{r}{$r$}
\psfrag{0}{$0$}
\psfrag{1}{$1$}
\psfrag{3}{$3$}
\psfrag{4}{$4$}
\psfrag{6}{$6$}
\psfrag{7}{$7$}
\psfrag{5}{$5$}
\centerline{
 \includegraphics[width=2in]{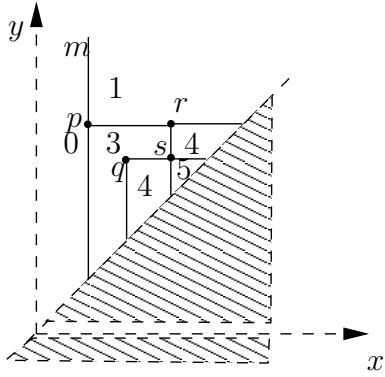}
} \caption{Cornerpoints of a reduced size function:  in this example, $p$, $q$ and
$r$ are the only proper cornerpoints, and have multiplicity equal to $2$ ($p$) and $1$ ($r,q$). The point $s$ is not a cornerpoint, since its multiplicity vanishes. The line $m$ is the only cornerpoint at infinity. }\label{Fig:value}
\end{figure}

As an example of cornerpoints in reduced size functions, in Fig.~\ref{Fig:value} we see that the proper cornerpoints are the points $p$,
$q$ and $r$ (with multiplicity $2$, $1$ and $1$, respectively). The
line $m$ is the only cornerpoint at infinity.

\par
\par

The importance of cornerpoints is revealed by the next  result, analogous to Prop.~10 of \cite{landi2001}, showing that cornerpoints,  with their multiplicities,
uniquely determine  reduced size functions.

The open (resp. closed) half-plane $\Delta^+$ (resp. $\bar
\Delta^+$) extended by the points at infinity of the kind $(k,\infty)$
will be denoted by $\Delta^*$ (resp. $\bar\Delta^*$), i.e.
$$\Delta^*:=\Delta^+\cup\{(k,\infty):k\in\br\}, \ \ \bar \Delta^*:=\bar \Delta^+\cup\{(k,\infty):k\in\br\}.$$

\begin{thm}{\rm (Representation Theorem)}
\label{value} For every  $({\bar x},{\bar y})\in\Delta^+$ we have
\begin{eqnarray}\label{reprthm}
\ell ^*_{({\M},\varphi )}({\bar x},{\bar y})=\sum _{ (x,y)\in\Delta^*\atop x\le {\bar x}, y>\bar y }\mu\big((x,y)\big).
\end{eqnarray}
\end{thm}

The equality~(\ref{reprthm})  can be checked in the example of Fig.~\ref{Fig:value}.
The points where the reduced size function  takes value $0$ are
exactly those for which there is no cornerpoint (either proper or at infinity)  lying to the left
and above them. Let us take a point  in the region of the domain
where the reduced size function takes the value  $3$. According to the
above theorem, the value of the reduced size function at that point must
be equal to $\mu(m)+\mu(p)=3$.

\begin{rem}\label{adv}{\em 
By comparing  Th.~\ref{value} and  the analogous result stated in
 Prop.~10 of \cite{landi2001}, one can observe that the former
is stated  more straightforwardly. As a consequence
of this simplification, all the statements  in this paper  that
follow from Th.~\ref{value} are less cumbersome than they would be if we applied size functions instead of reduced size functions.
This is the main
motivation for introducing the notion of reduced size function.}
\end{rem}

In order to make this paper self-contained, in the rest of this section  we report  all and only those results  about size functions that will be needed for proving our statements in the next sections, re-stating them in terms of reduced size functions. Proofs are omitted, since they are
completely analogous to those for classical size functions.

The following result, expressing a relation between two  reduced size
functions corresponding to  two spaces, $\M$ and $\N$, that can be
matched without changing the measuring functions more that $h$, is
analogous to Th.~3.2 in \cite{morftransf}.

\begin{prop}
\label{saltodiag} Let $(\M,\p)$ and $(\N,\s)$ be two size pairs.
If $f:\M\rightarrow \N$ is a  homeomorphism such that $\max_{P\in
\M}|\p(P)-\s(f(P))|\le h$, then for every $({\bar x},{\bar
y})\in\Delta^+$  we have
$$\rsfp({\bar x}-h,{\bar y}+h)\le \rsfn({\bar x},{\bar y}).$$
\end{prop}

The next proposition, analogous to Prop.~6 in \cite{landi2001}, gives some constraints on the presence of discontinuity points for reduced  size functions.

\begin{prop}
\label{constr} Let $(\M,\p)$ be a size pair. For every point
${\bar p}=({\bar x},{\bar y})\in \Delta^+$, a real number
$\epsilon>0$ exists such that  the  open set
$$W_\epsilon ({\bar p}):=\{(x,y)\in\br ^2: |{\bar x}-x|< \epsilon, |{\bar
y}-y|<\epsilon,x\ne {\bar x},y\ne {\bar y}\}$$ is contained in
$\Delta^+$, and  does not contain any discontinuity point  for
$\rsfp$.
 \end{prop}

The following analog of Prop.~8 and Cor.~4 in \cite{landi2001}, stating that
cornerpoints create discontinuity points spreading downwards and
towards the right to $\Delta$, also holds  for reduced size
functions.

\begin{prop}{\rm (Propagation of discontinuities)}
\label{corn=disc} If ${\bar p}=({\bar x},{\bar y})$ is a
proper cornerpoint for   $\rsfp$, then the following statements hold:

{\rm i)} If ${\bar x}\le x<{\bar y}$, then ${\bar y}$ is a
discontinuity point for $\rsfp(x,\cdot )$;

{\rm ii)} If ${\bar x}<y<{\bar y}$, then ${\bar x}$ is a
discontinuity point for $\rsfp(\cdot, y)$.

\noindent If ${\bar r}=({\bar x},{\infty})$ is the
 cornerpoint at infinity for   $\rsfp$, then the following statement holds:

{\rm iii)} If ${\bar x}< y$, then ${\bar x}$ is a
discontinuity point for $\rsfp(\cdot,y )$.
\end{prop}

The position of cornerpoints in $\Delta^+$ is related to the extrema  of the measuring function as the next proposition  states, immediately following from the definitions.

 \begin{prop}{\rm (Localization of cornerpoints)}
\label{corn=pos} If ${\bar p}=({\bar x},{\bar y})$ is a
proper cornerpoint for   $\rsfp$, then
$$\bar p\in\{(x,y)\in\br^2:\min\p\le x<y\le \max\p\}.$$

If ${\bar r}=({\bar x},\infty)$ is the
 cornerpoint at infinity for   $\rsfp$, then $\bar x=\min\p$.
\end{prop}

Prop.~\ref{corn=disc} and Prop.~\ref{corn=pos} imply that the number of cornerpoints is either finite or countably infinite. In fact, the following result can be proved,
analogous to  Cor.~3 in \cite{landi2001}.

\begin{prop}{\rm (Local finiteness of cornerpoints)}
\label{corncount} For each strictly positive real  number
$\epsilon$, reduced size functions have, at most, a finite number of
cornerpoints in $\{(x,y)\in\br^2:x+\epsilon<y\}$.
\end{prop}

\par
\begin{figure}
\psfrag{x}{$x$}
\psfrag{y}{$y$}
\psfrag{p}{$\varphi$}
\centerline{
 \includegraphics[height=5cm]{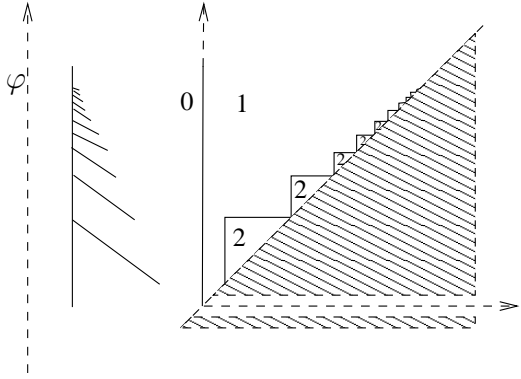}
} \caption{A reduced size function with cornerpoints accumulating
onto the diagonal: $\M$ is the space depicted on the left and $\p$
measures the height of each point. }\label{Fig:accumul}
\end{figure}
\par

Therefore, if the set of cornerpoints of a reduced size
function has an accumulation point, it necessarily belongs to the
diagonal $\Delta$.  An example of reduced size function with
cornerpoints accumulating onto the diagonal is shown in
Fig.~\ref{Fig:accumul}.

Moreover, this last proposition implies that in the  summation of Th.~\ref{value} (Representation Theorem), only  finitely many terms are different from zero.

\section{Matching distance}
In this section we define a matching distance between reduced size
functions. The idea is to compare reduced size functions by
measuring the cost of transporting the cornerpoints of one reduced
size function to those of the other one, with the property that
the longest of the transportations should be as short as possible.
Since, in general, the number of cornerpoints of the two reduced
size functions is different, we also enable the  cornerpoints
to be transported onto the points of $\Delta$ (in other words, we can ``destroy'' them). 

When the number of cornerpoints is
finite, the matching distance may be related to  the bottleneck transportation problem (cf., e.g.,
 \cite{Efrat}, \cite{Garfinkel}).   In our case, however, the number
of cornerpoints may be countably infinite, because of our loose assumption on the measuring function, that is only required to be continuous. Nevertheless, we prove
the existence of an optimal matching. Under  more tight assumptions on  the measuring function, the number of cornerpoints is  ensured to be finite and a bottleneck distance can be more straightforwardly defined. For example,   in \cite{CoEdHa07} a bottleneck distance for comparing Persistent Homology Groups is introduced under the assumption that the measuring functions are {\em tame}. We recall that a continuous function is tame if it has a finite number of homological critical values and the homology groups of the lower level sets it defines are finite dimensional. 

Although working with measuring functions that are continuous rather than tame involves working in an infinite dimensional space,  yielding many technical difficulties in the proof of our results (for instance, compare our Matching Stability Theorem \ref{main} with the analogous Bottleneck Stability Theorem for Persistence Diagrams in  \cite{CoEdHa07}), there are strong motivations for doing so. First of all, in real applications noise cannot be assumed to be tame, so that the perturbation of a tame function may happen to be not tame. In second place,  when working in the more general setting of measuring functions with  values in ${\mathbb R}^k$ instead of ${\mathbb R}$, it is important that the  set of functions is closed under the action of the $\max$ operator, as shown in \cite{CaDiFe07}, whereas the set of tame functions is not. Last but not least, working with continuous functions    
 allows us to relate the matching distance to the natural pseudo-distance, which is our final goal, without restricting the set of homeomorphism to those preserving the tameness property.

Of course the matching distance
is not the only metric between reduced size functions that we
could think of. Other metrics for size functions have been considered in the past
(\cite{york}, \cite{sdiego}). However,  the
matching distance is of particular interest since, as we shall see,
it allows for a connection with the natural pseudo-distance between
size pairs, furnishing the best possible lower bound. Moreover, it has already been experimentally tested
successfully in \cite{EVAM} and \cite{CeFeGi06}.

In order to introduce the matching distance between reduced size functions we need some new definitions. The following definition of representative sequence is introduced in order to manage the presence in a size function of infinitely many cornerpoints as well as that of their multiplicities. Moreover, it allows us to add to the set of cornerpoints a subset of points of the diagonal.  

\begin{defn}{\rm (Representative sequence)}
Let $\ell^*$ be a reduced size function. We shall call {\em
representative sequence  for $\ell^*$} any sequence of points
$a:\bn\rightarrow {\bar \Delta^*}$, (briefly denoted by $(a_i)$),
with the following properties:
\begin{enumerate}
\item $a_0$ is the cornerpoint at infinity for $\ell^*$;
\item For each $i>0$, either $a_i$ is a proper cornerpoint for $\ell^*$, or $a_i$ belongs to $\Delta$;
\item If $p$ is a proper cornerpoint for $\ell^*$ with multiplicity $\mu(p)$, then the cardinality of the set $\{i\in \bn:a_i=p\}$
 is equal to  $\mu(p)$;
\item The set of indexes  for which $a_i$ is in $ \Delta$ is countably infinite.
\end{enumerate}
\end{defn}

 We now consider the following pseudo-distance $d$ on
 $\bar \Delta^*$ in order to assign a cost to each deformation of reduced size functions:
$$d\left(\left(x,y\right),\left(x',y'\right)\right):=
\min\left\{\max\left\{|x-x'|,|y-y'|\right\},\max \left\{\frac{y-x}{2},\frac{y'-x'}{2}\right\}\right\},$$
with the convention about $\infty$ that $\infty-y=y-\infty=\infty$ for $y\ne \infty$, $\infty-\infty=0$, $\frac{\infty}{2}=\infty$,
$|\infty|=\infty$, $\min\{\infty,c\}=c$, $\max\{\infty,c\}=\infty$.

In other words, the pseudo-distance $d$ between two points $p$ and
$p'$  compares the cost of moving $p$ to $p'$
and the cost of moving $p$ and $p'$ onto the diagonal and takes the smaller. Costs
are computed   using the distance  induced by the
$\max$-norm. In particular, the pseudo-distance $d$ between two
points $p$ and $p'$ on the diagonal is always $0$; the
pseudo-distance $d$ between two points $p$ and $p'$, with $p$
above the diagonal and $p'$  on the diagonal, is equal to the
distance, induced by the $\max$-norm, between $p$ and the
diagonal. Points at infinity have a finite distance only to other points at infinity, and their distance depends on their abscissas.

Therefore,  $d(p,p')$ can  be considered  a measure of  the minimum
of the costs of moving $p$ to $p'$ along two different paths (i.e.
the path that takes $p$  directly to $p'$ and the path that passes
through $\Delta$). This observation easily yields that $d$ is
actually a pseudo-distance.

\begin{rem}{\em
It is useful to observe what disks  induced by the
pseudo-distance $d$ look like. For $r>0$, the usual notation
$B(p,r)$ will denote the open disk $\{p'\in  \bar
\Delta^+:d(p,p')<r\}$. Thus, if $p$ is a proper point with coordinates $(x,y)$ and
$y-x\ge 2r$ (that is, $d(p,\Delta)\ge r$), then $B(p,r)$ is the
open square centered at $p$ with sides of length $2r$ parallel to the axes. Whereas, if
$p$ has coordinates $(x,y)$ with $y-x< 2r$ (that is, $d(p,\Delta)<
r$), then $B(p,r)$ is the union of the open square, centered at $p$,
with sides of length $2r$ parallel to the axes, with the stripe $\{(x,y)\in \br^2:  0\le
y-x<2r\}$, intersected with $\bar \Delta^+$(see also
Fig.~\ref{Fig:1}). If $p=(x,\infty)$ is a point at infinity then $B(p,r)=\{(x',\infty)\in\Delta^*:|x-x'|<r\}$.}

\par
\begin{figure}[h]
\centerline{\hbox{
 \includegraphics[width=2.5in]{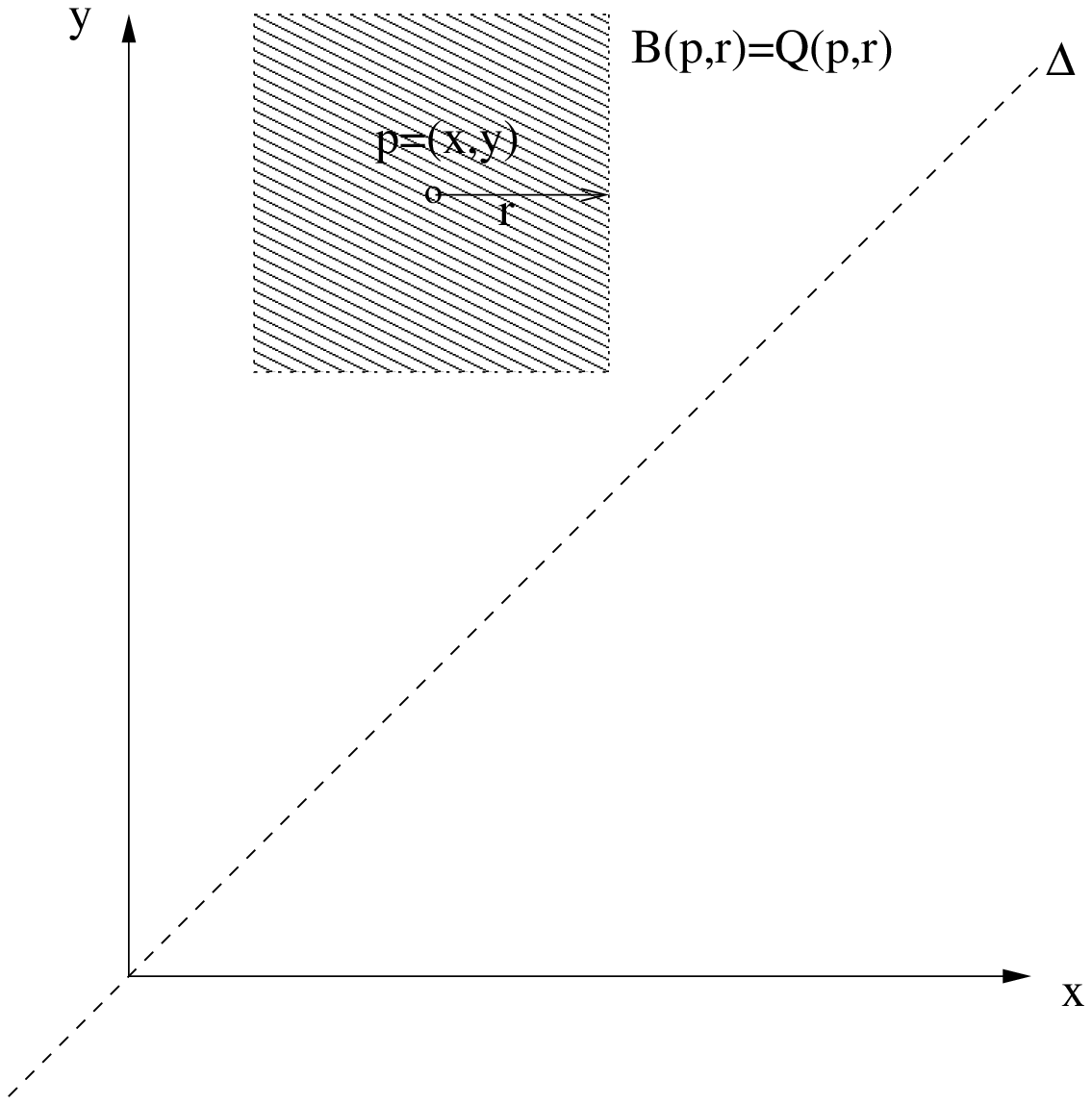}
 \includegraphics[width=2.6in]{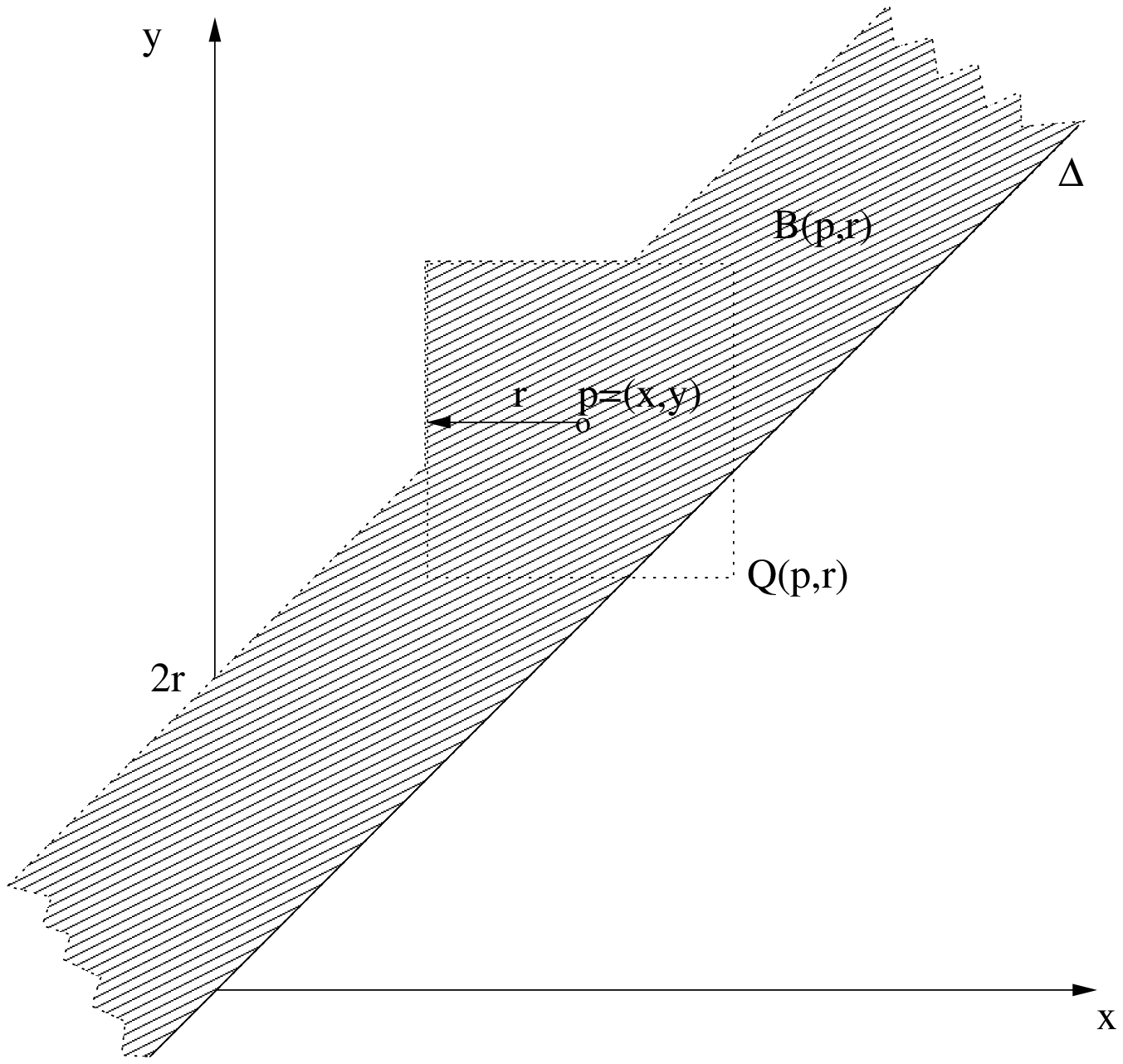}
}} \caption{Disks induced by the pseudo-metric $d$ (shaded). Left:
$d(p,\Delta)\ge r$. Right: $d(p,\Delta)<r$.}\label{Fig:1}
\end{figure}
\par
\end{rem}

In what follows, the notation $Q(p,r)$ will refer to the open square
centered at the proper point $p$, with sides of length $2r$ parallel to the axes (that is, the open disk centered at
$p$ with radius $r$, induced by the $\max$-norm). Also, when $r>0$, ${\bar
Q}(p,r)$ will refer to the closure of $Q(p,r)$ in the usual Euclidean topology, while ${\bar Q}(p,0):=\{p\}$.

\begin{defn}{\rm (Matching distance)}
Let $\ell^*_1$ and $\ell^*_2$ be two reduced size functions. If
$(a_i)$ and $(b_i)$ are two representative sequences for
$\ell^*_1$ and $\ell^*_2$ respectively, then the {\em matching
distance} between $\ell^*_1$ and $\ell^*_2$ is the number
$$d_{match}(\ell^*_1,\ell^*_2):=\inf_\sigma\sup_i d(a_i,b_{\sigma(i)}),$$
where $i$ varies in $\bn$ and $\sigma$ varies among all the bijections from $\bn$ to $\bn$.
\end{defn}

\begin {figure}[h]
 \begin{center}\begin{tabular}{cc}
\begin{tabular}{cc}
\includegraphics[height=3cm]{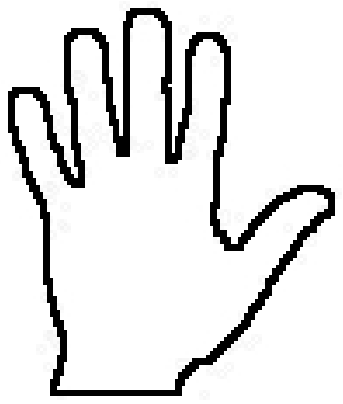}&
\includegraphics[height=4cm]{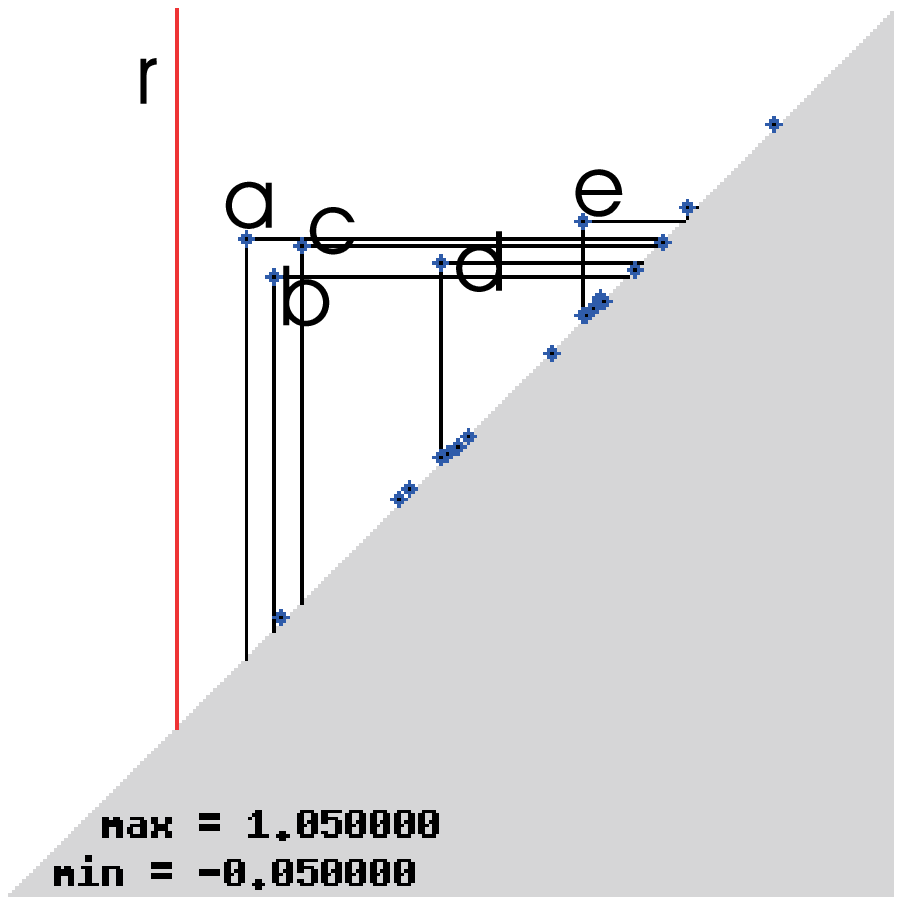} \\
\includegraphics[height=3cm]{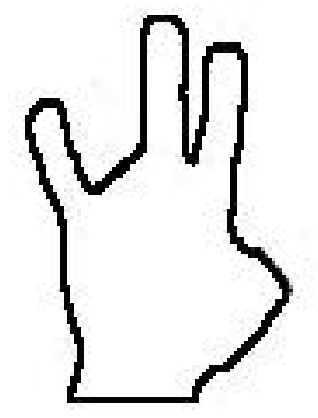}&
\includegraphics[height=4cm]{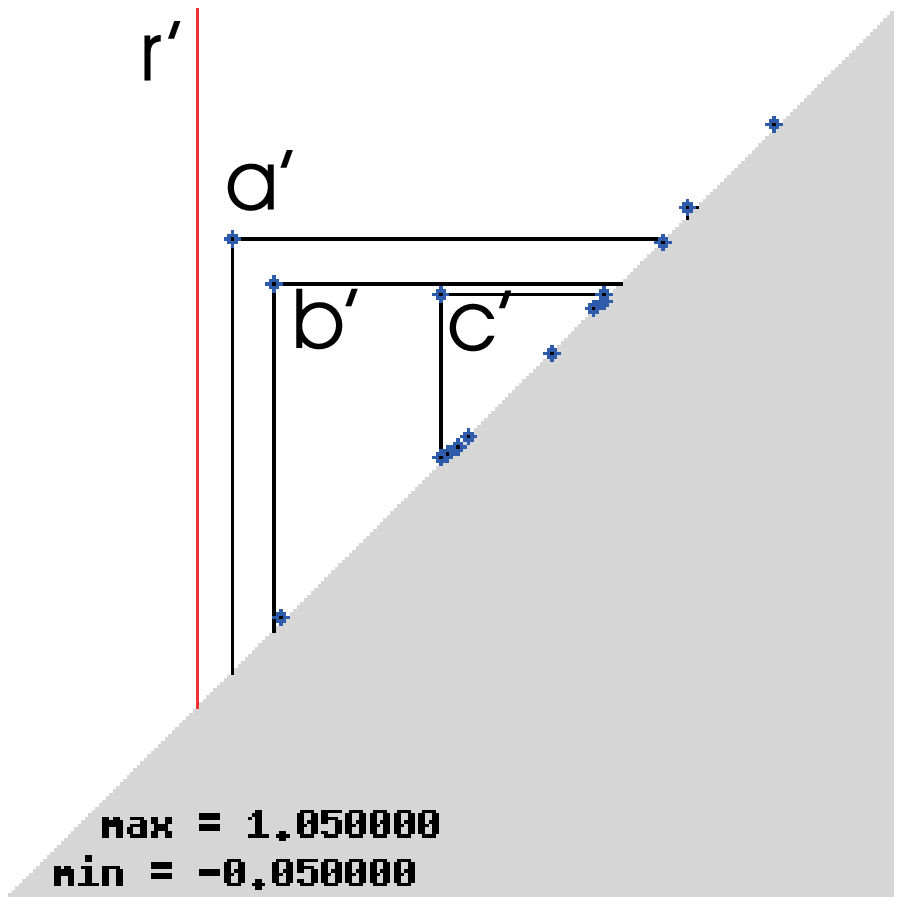}
\end{tabular} &
\begin{tabular}{c}
\ \\
\includegraphics[height=5cm]{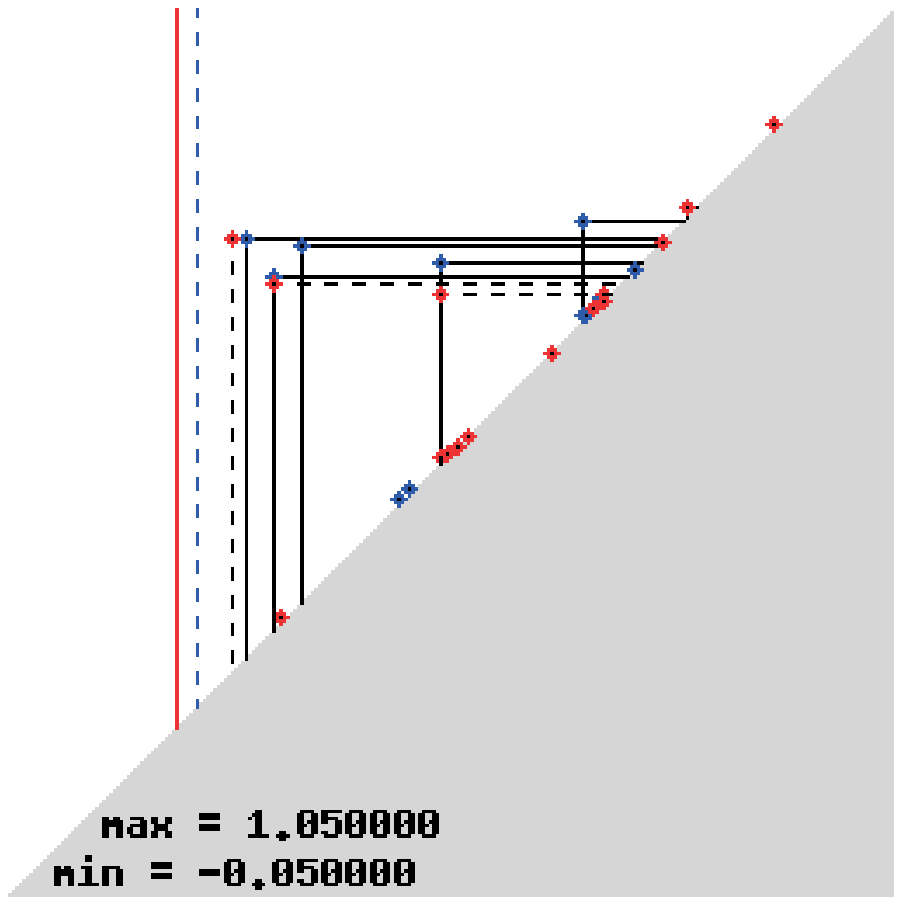}
\end{tabular}
\end{tabular}
\end{center}
\caption{ Left: Two curves. Center: Their reduced size
functions with respect to the measuring function distance from the center of the image. Right: The
superimposition of the two reduced size functions.}
\label{esempiomatching}
 \end {figure}

In order to illustrate this definition, let us consider
Fig.~\ref{esempiomatching}. Given two curves, their reduced size
functions with respect to the measuring function distance from the
center of the image are calculated. One sees that the top reduced size
function has many cornerpoints close to the diagonal in addition to the
cornerpoints  $r$, $a$, $b$, $c$, $d$, $e$. Analogously, the
bottom reduced size function has many cornerpoints close to the diagonal
in addition to the cornerpoints  $r'$, $a'$, $b'$, $c'$. Cornerpoints close to
the diagonal are generated by noise and discretization. The
superimposition of the two reduced size functions shows that an optimal
matching is given by $r\to r'$, $a\to a'$, $b\to b'$, $c\to c'$,
$d\to \Delta$, $e\to\Delta$, and all the other cornerpoints sent
to $\Delta$. Sending cornerpoints to points of $\Delta$
corresponds to the annihilation of cornerpoints.
 Since the matching $c\to c'$ is the one that achieves the
maximum cost in the $\max$-norm, the matching distance is equal to
the distance between $c$ and $c'$ (with respect to the $\max$-norm).

\par
\par

\begin{prop}\label{dist}
$d_{match}$ is a distance between reduced size functions.
\end{prop}

\begin{pf}
It is easy to see that this definition is independent from the
choice of the representative sequences of points for $\ell^*_1$
and $\ell^*_2$. In fact, if $(a_i)$ and $({\hat a}_i)$ are
representative sequences for the same reduced size function
$\ell^*$, a bijection $\hat\sigma:\bn\rightarrow \bn$ exists such
that $d({\hat a}_i,a_{{\hat\sigma}(i)})=0$ for every index $i$.

Furthermore, we have that $d_{match}(\rsfp,\rsfn)<+\infty$, for any two size pairs $(\M,\p)$ and $(\N,\s)$. Indeed, for any
bijection $\sigma:\bn\rightarrow \bn$ such that $\sigma(0)=0$, it holds $\sup_i d(a_i,b_{\sigma(i)})<+\infty$, because
of Prop.~\ref{corn=pos} (Localization of cornerpoints).

Finally, by recalling  that reduced size functions are
uniquely determined by their cornerpoints with  multiplicities (Representation Theorem) and by using Prop.~\ref{corncount} (Local finiteness of cornerpoints), one can easily see that
$d_{match}$ verifies all the properties of  a distance.
\qed\end{pf}

We will show in Th.~\ref{lem:minmax} that the $\inf$ and the
$\sup$ in the definition of matching distance are actually
attained, that is
$d_{match}(\ell^*_1,\ell^*_2)=\min_\sigma\max_i
d(a_i,b_{\sigma(i)})$. In other words,  an optimal
matching always exists.

\section{Stability of the matching distance}

In this section we shall prove  that if $\p$ and $\s$ are  two
measuring functions on $\M$ whose difference on the points of $\M$
is controlled by $\epsilon$
 (namely
$\max_{P\in \M}|\p(P)-\s(P)|\le \epsilon$), then the matching
distance  between $\rsfp$ and $\rsfs$ is also controlled by
$\epsilon$ (namely $d_{match}(\rsfp,\rsfs)\le\epsilon$).

For the sake of clarity, we will now give  a sketch of the proof that
will lead to this result, stated in Th.~\ref{main}. We begin by
proving that each cornerpoint of $\rsfp$ with multiplicity $m$
admits  a small neighborhood, where we find exactly $m$
cornerpoints (counted with multiplicities) for $\rsfs$, provided
that on $\M$ the functions $\p$ and $\s$ take  close enough values
(Prop.~\ref{lem:1}). Next, this local result is extended to a
global result by considering the convex combination
$\Phi_t=\frac{t}{\epsilon}\s+\frac{\epsilon-t}{\epsilon}\p$  of
$\p$ and $\s$. Following the paths traced by the cornerpoints of
$\ell^*_{(\M,\Phi_t)}$ as $t$ varies in $[0,\epsilon]$, in
Prop.~\ref{prop:2} we show that, along these paths, the displacement of the cornerpoints
is not greater than  $\epsilon$ (displacements are measured using the
distance $d$, and cornerpoints are counted with their multiplicities).
Thus we are able to construct an injection $f$, from the set of the
cornerpoints of $\rsfp$  to the set of the cornerpoints of $\rsfs$
(extended to a countable subset of the diagonal), that moves points less
than $\epsilon$ (Prop.~\ref{prop:3}). Repeating the same argument
backwards,  we construct  an injection $g$ from the set of the
cornerpoints of $\rsfs$ to the set of the cornerpoints of $\rsfp$
(extended to a countable subset of the diagonal) that moves points less than
$\epsilon$. By using  the Cantor-Bernstein theorem, we prove that  there exists a bijection
from the set of the cornerpoints of $\rsfs$ to the set of  the cornerpoints of $\rsfp$
(both the sets extended to countable subsets of the diagonal) that moves points less than
$\epsilon$. This will be sufficient to conclude the proof. Once
again, we  recall that in the proof we have just  outlined,
cornerpoints are always counted with their  multiplicities.

We first prove that the number of proper cornerpoints contained in a
sufficiently small square can be computed in terms of jumps of
reduced size functions.

\begin{prop}\label{lem:0}
Let $(\M,\p)$ be a size pair. Let ${\bar p}=({\bar x},{\bar y})\in \Delta ^+$ and let $\eta>0$ be such that  ${\bar x}+\eta<{\bar y}-\eta$. Also let
$a=({\bar x}+\eta,{\bar y}-\eta)$, $b=({\bar x}-\eta,{\bar y}-\eta)$,
$c=({\bar x}+\eta,{\bar y}+\eta)$, $e=({\bar x}-\eta,{\bar y}+\eta)$. Then
$$\rsfp(a)-\rsfp(b)-\rsfp(c)+\rsfp(e)$$
is equal to the number of (proper) cornerpoints for $\rsfp$, counted with
their multiplicities, contained  in the semi-open square  $\hat Q_\eta$, with
vertices at $a,b,c,e$, given by
$$\hat Q_\eta := \{(x,y)\in \Delta^+:  {\bar x}-\eta<x\le {\bar x}+\eta, \ {\bar y}-\eta< y\le {\bar y}+\eta\}.$$
\end{prop}

\begin{pf}
It  easily follows from the Representation Theorem (Th.~\ref{value}). \qed\end{pf}

We now show that, locally, small changes in the measuring functions produce small displacements of the existing proper cornerpoints and create no new cornerpoints.

\begin{prop}{\rm (Local constancy of multiplicity)}\label{lem:1}
Let $(\M,\p)$ be a size pair and let ${\bar p}=({\bar x},{\bar y})$ be a point  in $\Delta^+$,
with multiplicity $\mu ({\bar p})$ for $\rsfp$ (possibly $\mu(\bar p)=0$).
Then there is a real number $\bar\eta>0$ such that, for any real number $\eta$ with $0\le \eta\le\bar\eta$, and
for any  measuring function $\s:\M\rightarrow \br$ with
 $\max_{P\in \M}|\p(P)-\s(P)|\le\eta$, the reduced size function $\rsfs$ has exactly $\mu({\bar p})$ (proper) cornerpoints  (counted with
their multiplicities) in the  closed square ${\bar Q}({\bar
p},\eta)$, centered at $\bar p$ with side $2\eta$.
\end{prop}

\begin{pf}
By Prop.~\ref{constr}, a sufficiently small real number $\epsilon >0$
exists
such that  the   set
$$W_\epsilon({\bar p})=\{(x,y)\in\br ^2: |{\bar x}-x|< \epsilon, |{\bar
y}-y|<\epsilon,x\ne {\bar x},y\ne {\bar y}\}$$
is contained in $\Delta^+$ (i.e. $\bar x+\epsilon\le \bar y-\epsilon$), and does not contain any discontinuity point  for $\ell
^*_{(\M,\varphi )}$. Prop.~\ref{corn=disc} (Propagation of discontinuities) implies that $\bar p$ is the only cornerpoint in $Q(\bar p,\epsilon)$.

\par
\begin{figure}
\psfrag{a}{$a$}
\psfrag{b}{$b$}
\psfrag{c}{$c$}
\psfrag{e}{$e$}
\psfrag{w}{$W_\epsilon(\bar p)$}
\psfrag{q}{$\hat Q_{\eta+\delta}$}
\psfrag{p}{$\bar p=(\bar x,\bar y)$}
\psfrag{x}{$(\bar x+\delta,\bar y-\delta)$}
\psfrag{y}{$\ (\bar x+2\eta+\delta,\bar y-2\eta-\delta)$}
\psfrag{h}{$\epsilon$}
\centerline{\hbox{
 \includegraphics[width=2.5in]{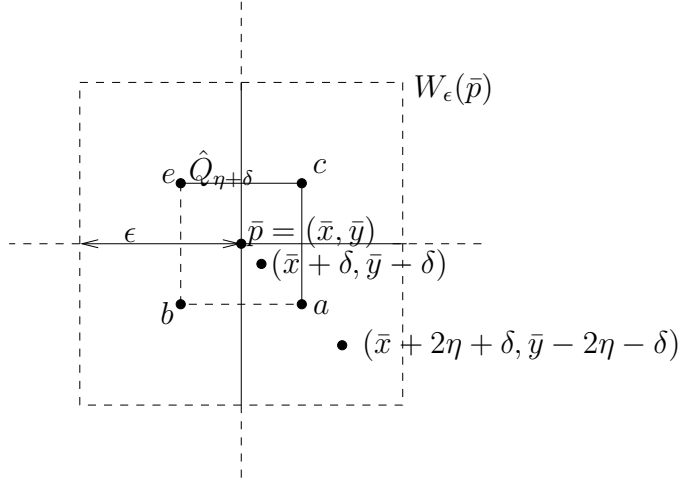}
}} \caption{The sets $W_\epsilon({\bar p})$ and $\hat Q_{\eta+\delta}$  used in the proof of
Prop.~\ref{lem:1}.}\label{Fig:wpe}
\end{figure}
\par

Let $\bar\eta$ be any real number such that $0<\bar\eta<\epsilon/2$.  For each real number $\eta$ with $0\le \eta\le \bar\eta$, let us  take a sufficiently small positive real number $\delta$ with $\delta<\eta$ and $2\eta+\delta<\epsilon$, so that $\bar x+2\eta+\delta <\bar y-2\eta-\delta$.

We define $a=({\bar x}+\eta+\delta,{\bar y}-\eta-\delta)$, $b=({\bar x}-\eta-\delta,{\bar y}-\eta-\delta)$,
$c=({\bar x}+\eta+\delta,{\bar y}+\eta+\delta)$, $e=({\bar x}-\eta-\delta,{\bar y}+\eta+\delta).$

If $\s:\M\rightarrow \br$ is a measuring function such that $\max_{P\in\M}|\p(P)-\s(P)|\le \eta$, then by applying Prop.~\ref{saltodiag} twice,
$$\rsfp(\bar x+\delta,\bar y-\delta)\le \rsfs(\bar x+\eta+\delta,\bar y-\eta-\delta)\le \rsfp(\bar x+2\eta+\delta,\bar y-2\eta-\delta).$$
Since $\rsfp$ is constant in each connected component of $W_\epsilon(\bar p)$, we have that
$$\rsfp(\bar x+\delta,\bar y-\delta)=\rsfp(a)=\rsfp(\bar x+2\eta+\delta,\bar y-2\eta-\delta),$$
implying $\rsfp(a)=\rsfs(a)$. Analogously,  $\rsfp(b)=\rsfs(b)$, $\rsfp(c)=\rsfs(c)$, $\rsfp(e)=\rsfs(e)$.
 Hence, $\mu(\bar p)$, i.e.\/ the multiplicity of  $\bar p$ for $\rsfp$, equals
$$\rsfs(a)-\rsfs(b)-\rsfs(c)+\rsfs(e).$$
 By Prop.~\ref{lem:0}, we obtain that $\mu(\bar p)$ is equal to the number of cornerpoints for $\rsfs$ contained in the semi-open square with vertices $a$, $b$, $c$, $e$ given by
$$\hat Q_{\eta+\delta}=\{(x,y)\in\Delta^+:\bar x-\eta-\delta<x\le \bar x+\eta+\delta, \bar y-\eta-\delta<y\le \bar y+\eta+\delta\}.$$
 This is true for any sufficiently small $\delta>0$. Therefore, $\mu(\bar p)$ is equal to  the number of cornerpoints for $\rsfs$ contained in the intersection $\bigcap_{\delta>0}\hat Q_{\eta+\delta}$. It follows that $\mu(\bar p)$  amounts to the number of cornerpoints for $\rsfs$ contained in the closed square $\bar Q(\bar p,\eta)$.
\qed\end{pf}

The following result states  that if two measuring functions $\p$ and
$\s$  differ less than $\epsilon$ in the $L_\infty$-norm, then it
is possible to  match some finite sets of proper cornerpoints of $\rsfp$ to
proper cornerpoints of $\rsfs$, with a motion smaller than $\epsilon$.

\begin{prop}\label{prop:2}
Let $\epsilon\ge 0$ be a real number and let $(\M,\p)$ and
$(\M,\s)$ be two size pairs such that $\max_{P\in
\M}|\p(P)-\s(P)|\le\epsilon$. Then, for any  finite set $K$ of
proper cornerpoints   for $\rsfp$ with $d(K,\Delta)>\epsilon$, there exist
$(a_i)$ and $(b_i)$ representative sequences for $\rsfp$ and
$\rsfs$ respectively, such that  $d(a_i,b_{i})\le\epsilon$
for each $i$ with $a_i\in K$.
\end{prop}

\begin{pf}
The claim is trivial for $\epsilon=0$, so let us assume $\epsilon>0$.

Let $\Phi_t=\frac{t}{\epsilon}\s+\frac{\epsilon-t}{\epsilon}\p$
with $t\in [0,\epsilon]$. Then, for every $t,t'\in [0,\epsilon]$,
we have  $\max_{P\in \M}|\Phi_t(P)-\Phi_{t'}(P)|\le|t-t'|$.

 Let $K=\{p_1,\ldots ,p_k\}$, let $m_j$ be the multiplicity  of $p_j$, for  $j=1, \ldots ,k$,
 and $m=\sum_{j=1}^km_j$.
Then we can easily construct a representative sequence of points
$(a_i)$ for $\rsfp$, such that
\begin{eqnarray*}
a_1=p_1, \ldots , a_{m_1}=p_1,  a_{m_1+1}=p_2,
\ldots , a_{m_1+m_2}=p_2, \ldots ,\\
a_{m_1+\ldots+m_{k-1}}=p_{k-1},a_{m_1+\ldots+m_{k-1}+1}=p_{k},\ldots , a_{m}=p_k.
\end{eqnarray*}
Now we will consider the set $A$ defined as
$$\{\delta\in[0,\epsilon]:\hbox{$\exists (a_i^\delta)$ representative sequence for $\ell^*_{(\M,\Phi_\delta)}$
s.t. $d(a_i,a_{i}^\delta)\le\delta$, $\forall a_i\in K$}\}.$$
 In other words, if we think of  the variation of $t$ as  the flow of time, $A$ is the set of  instants  $\delta$  for which
the cornerpoints in $K$ move less than $\delta$ itself, when the
homotopy $\Phi_t$ is applied to the measuring function $\p$.

$A$ is non-empty, since $0\in A$. Let us set $\bar\delta=\sup A$ and show that
$\bar\delta\in A$. Indeed, let $(\delta_n)$ be a sequence of
numbers of $A$ converging to $\bar\delta$. Since $\delta_n\in A$,
for each $n$ there is a representative sequence $(a_i^{\delta_n})$
for $\ell^*_{(\M,\Phi_{\delta_n})}$ with
$d(a_i,a_{i}^{\delta_n})\le\delta_n$, for each $i$ such that
$a_i\in K$. Since $\delta_n\le \epsilon$,
$d(a_i,a_i^{\delta_n})\le\epsilon$ for any $i$ and any $n$. Thus, recalling that $d(K,\Delta)>\epsilon$,
for each $i$ such that $a_i\in K$, it holds that $a_i^{\delta_n}\in \bar
Q(a_i,\epsilon)$ for any $n$. Hence,
for each $i$ with $a_i\in K$, possibly by extracting a convergent
subsequence, we can define
$a_i^{\bar\delta}=\lim_na_i^{\delta_n}$. We have
$d(a_i,a_i^{\bar\delta})\le\bar\delta$. Moreover, by
Prop.~\ref{lem:1} (Local constancy of multiplicity), $a_i^{\bar\delta}$ is a cornerpoint for
$\ell^*_{(\M,\Phi_{\bar\delta})}$. Also, if  $r$
indexes $j_1,\ldots, j_r$ exist, such
 that $a_{j_1},\ldots, a_{j_r}\in K$ and $a_{j_1}^{\bar\delta}=\cdots = a_{j_r}^{\bar\delta}=:q$, then the
 multiplicity  of $q$ for $\ell^*_{(\M,\Phi_{\bar\delta})}$ is
 not smaller than $r$. Indeed, since $\delta_n\to\bar\delta$, for each arbitrarily small
 $\eta>0$ and  for any sufficiently great $n$,  the square
 $\bar Q(q,\eta)$ contains at least $r$
 cornerpoints for $\ell^*_{(\M,\Phi_{\delta_n})}$, counted with
 their multiplicities. But Prop.~\ref{lem:1}  implies that, for each sufficiently small $\eta$, $\bar Q(q,\eta)$
 contains  exactly as many  cornerpoints for
 $\ell^*_{(\M,\Phi_{\delta_n})}$ as the multiplicity of $q$ with respect to $\ell^*_{(\M,\Phi_{\bar \delta})}$, if $|\delta_n-\bar\delta|\le \eta$.
 Therefore, the multiplicity of $q$ for $\ell^*_{(\M,\Phi_{\bar \delta})}$ is greater than, or equal to, $ r$.

The previous reasoning allows us to claim that if a cornerpoint $q$ occurs $r$ times in the sequence $(a_1^{\bar\delta},\ldots, a_{m}^{\bar\delta})$, then the multiplicity of $q$ for $\ell^*_{(\M,\Phi_{\bar\delta})}$ is at least $r$.

 In order to conclude  that $\bar\delta\in A$, it is now
 sufficient to observe that $(a_1^{\bar\delta},\ldots, a_{m}^{\bar\delta})$ is
  easily extensible to a representative sequence for
 $\ell^*_{(\M,\Phi_{\bar\delta})}$, simply by setting $a_0$ equal to  the cornerpoint at infinity of $\ell^*_{(\M,\Phi_{\bar\delta})}$, and by continuing the sequence with the remaining proper cornerpoints of $\ell^*_{(\M,\Phi_{\bar\delta})}$ and with a countable collection of points of $\Delta$. So we have proved that $\sup A$ is attained in $A$.

We end the proof by showing that $\max A=\epsilon$. In fact, if
$\bar\delta<\epsilon$, by using Prop.~\ref{lem:1} once again, it is not
difficult to show that there exists $\eta>0$, with $\bar \delta+\eta
<\epsilon$, and a representative sequence
$(a_i^{\bar\delta+\eta})$ for
$\ell^*_{(\M,\Phi_{\bar\delta+\eta})}$, such that
$d(a_i^{\bar\delta},a_i^{\bar\delta+\eta})\le \eta$ for  $1\le i\le
m$. Hence, by the triangular inequality,
$d(a_i,a_i^{\bar\delta+\eta})\le \bar\delta+\eta$ for  $1\le i\le
m$, implying that $\bar\delta+\eta\in A$. This would contradict
the fact that $\bar\delta=\max A$. Therefore, $\epsilon=\max A$, and so $\epsilon\in A$.
\qed\end{pf}

We now give a  result stating that if two measuring functions $\p$ and
$\s$  differ less than $\epsilon$ in the $L_\infty$-norm, then  the cornerpoints at infinity have a distance smaller than $\epsilon$.

\begin{prop}\label{displinf}
Let $\epsilon\ge 0$ be a real number and let $(\M,\p)$ and
$(\M,\s)$ be two size pairs such that $\max_{P\in
\M}|\p(P)-\s(P)|\le\epsilon$. Then, for each $(a_i)$ and $(b_i)$ representative sequences for $\rsfp$ and
$\rsfs$, respectively, it holds  that $d(a_0,b_{0})\le\epsilon$.
\end{prop}

\begin{pf}
By Prop.~\ref{corn=pos} (Localization of cornerpoints), $d(a_0,b_{0})=|\min\p-\min\s|$. Let $\min\p=\p(P_\p)$ and $\min\s=\s(P_\s)$, with $P_\p,P_\s\in\M$.
Since $\max_{P\in
\M}|\p(P)-\s(P)|\le\epsilon$, then
$$ \s(P_\s)\ge \p(P_\s)-\epsilon, \ \ \p(P_\p)\ge \s(P_\p)-\epsilon.$$
 By contradiction, let us assume that $d(a_0,b_0)>\epsilon$, that is,
$|\p(P_\p)-\s(P_\s)|>\epsilon$. So either $\p(P_\p)< \s(P_\s)-\epsilon$ or $\s(P_\s)< \p(P_\p)-\epsilon$.  In the first case,
$$\s(P_\p)-\epsilon\le \p(P_\p)< \s(P_\s)-\epsilon,$$
in the latter case,
$$\p(P_\s)-\epsilon\le \s(P_\s)< \p(P_\p)-\epsilon.$$
Hence we would conclude that either $P_\s$ is not a minimum point for $\s$ or $P_\p$ is not a minimum point for $\p$.
In both cases we get a contradiction.
\qed\end{pf}

Now we prove that it is possible to injectively match all  the
cornerpoints of $\rsfp$ to those of $\rsfs$ with a maximum motion
not greater than the $L_\infty$-distance between $\p$ and $\s$.

\begin{prop}\label{prop:3}
Let $\epsilon\ge 0$ be a real number and let $(\M,\p)$ and
$(\M,\s)$ be two size pairs such that
 $\max_{P\in \M}|\p(P)-\s(P)|\le\epsilon$.
 Then there exist $(a_i)$ and $(b_i)$  representative
sequences  for $\rsfp$ and $\rsfs$ respectively,  and an injection
$f:\bn\rightarrow\bn$  such that
$d(a_i,b_{f(i)})\le\epsilon$.
\end{prop}

\begin{pf}
The claim is trivial if $\epsilon=0$, so let us assume $\epsilon>0$.

Let $H$ be the set of all the cornerpoints (both proper and at
infinity) for $\rsfp$. We can write $H=K_1\cup K_2$, where
$K_1:=\{ p\in H: d(p,\Delta)> \epsilon\}$ and  $K_2:=\{ p\in H:
d(p,\Delta)\le \epsilon\}$.
The cardinality of $K_1$ is finite, according to  Prop.~\ref{corncount} (Local finiteness of cornerpoints).
Therefore, Prop.~\ref{prop:2} and Prop.~\ref{displinf} imply that
there exist $(a_i)$ and $(b_i)$ representative sequences of points
for $\rsfp$ and $\rsfs$ respectively,  such that
$d(a_i,b_{i})\le\epsilon$ for each $a_i\in K_1$ (it follows
that if $a_i\in K_1$ then, necessarily, $b_i\notin\Delta$).

Now let us write  $\bn$  as the disjoint union of the sets $I_1$
and $I_2$, where we set $i\in I_1$, if $a_i\in K_1$, and $i\in
I_2$, if $a_i\in K_2\cup\Delta$. We observe that, by the
definition of a representative sequence, there is a countably
infinite collection of indices $j$ with $b_j$ contained in
$\Delta$.  Thus, there is an injection $\beta: I_2\rightarrow \bn$
such that $b_{\beta(i)}\in\Delta$.

We define $f$ by setting $f(i)=i$ for $i\in I_1$ and
$f(i)=\beta(i)$ for $i\in I_2$. By construction, $f$ is injective
and $d(a_i,b_{f(i)})\le\epsilon$ for every $i$.
\qed\end{pf}

We recall the well-known
Cantor-Bernstein theorem (cf. \cite{Kuratowski}), which will be useful
later.

\begin{thm}{\rm (Cantor-Bernstein Theorem)}\label{cantor}
Let  $A$ and $B$ be two sets. If two injections $f:A\rightarrow B$
and $g:B\rightarrow A$ exist, then there is a bijection
$l:A\rightarrow B$. Furthermore, we can assume that the equality
$l(a)=b$ implies that either $f(a)=b$ or $g(b)=a$ (or both).
\end{thm}

We are now ready to prove a key result of this paper. We shall use this result
in the next section in order to prove that the matching distance between
reduced
size functions furnishes a lower bound for the natural
pseudo-distance between size pairs. Nevertheless, this result if meaningful by
itself, in that it guarantees the computational stability of the matching
distance between reduced size functions.

\begin{thm}{\rm (Matching Stability Theorem)}\label{main}
Let $(\M,\p)$ be a size pair. For every real number $\epsilon\ge0$ and for
every
measuring function $\s:\M\rightarrow \br$, such that
$\max_{P\in \M}|\p(P)-\s(P)|\le\epsilon$,   the matching distance between
$\rsfp$ and $\rsfs$ is smaller than or equal to $\epsilon$.
\end{thm}

\begin{pf}
Prop.~\ref{prop:3} implies that there exist $(a_i)$ and $(b_i)$
representative sequences  for $\rsfp$ and $\rsfs$ respectively,
and an injection $f:\bn\rightarrow\bn$  such that
$d(a_i,b_{f(i)})\le\epsilon$. Analogously, there exist $(a'_i)$
and $(b'_i)$  representative sequences  for $\rsfp$ and $\rsfs$
respectively,  and there is an injection $g':\bn\rightarrow \bn$,
such that  $d(b'_i,a'_{g'(i)})\le\epsilon$ for every
index $i$. Hence another injection $g:\bn\rightarrow \bn$ exists,
such that $d(b_i,a_{g(i)})\le\epsilon$ for every index $i$. Then
the claim follows from the Cantor-Bernstein Theorem, by setting
$A=B=\mathbb{N}$.
\qed\end{pf}

\section{The connection between the matching distance and the natural pseudo-distance}

We recall that, given two size
pairs $(\M,\p)$ and $(\N,\s)$, with $\M$ and $\N$ homeomorphic,  a measure of their shape
dissimilarity  is given by the
natural pseudo-distance.

As a corollary of the Matching Stability Theorem (Th.~\ref{main}) we obtain the following
Th.~\ref{nsd}, stating that the matching distance between reduced
size functions furnishes a lower bound for the natural
pseudo-distance between size pairs.

\begin{defn}
The natural pseudo-distance between two size
pairs $(\M,\p)$ and $(\N,\s)$ with $\M$ and $\N$ homeomorphic is the number
$$\inf_{h}\max _{P\in
\M}|\p(P)-\s(h(P))|,$$
where $h$ varies in the set $H(\M,\N)$ of all the homeomorphisms between $\M$ and $\N$.
\end{defn}

We point out that the natural pseudo-distance is not a distance because it can vanish on two non-equal size pairs. However, it is symmetric, satisfies the triangular inequality, and vanishes on two equal size pairs.

\begin{rem}{\em
We point out that an alternative definition of the dissimilarity
measure between size pairs, based on the integral of the change of
the measuring functions, rather than on the $\max$, may present some
drawbacks.

For example, let us consider the following size pairs $(\M,\p)$,
$(\N,\s)$, $(\N,\chi)$, where $\M$ is a circle of radius $2$,
$\N$ is a circle of radius $1$, and the measuring functions are
constant functions given by $\p\equiv 1$, $\s\equiv 1$,
$\chi\equiv 2$. Let $\mu$ and $\nu$ respectively denote  the $1$-dimensional measures
induced by the usual embeddings of $\M$ and $\N$ in the Euclidean
plane. By setting, for any homeomorphism $f:\M\rightarrow \N$,
$$\hat \Theta(f)=\int_{\M}\left|\varphi-\psi\circ f \right|d\mu,$$
 we have $\hat \Theta(f)=0$.
For any homeomorphism $g:\N\rightarrow \N$, we have
$$\hat
\Theta(g)=\int_{\N}\left|\s-\chi\circ g \right|d\nu=2\pi.$$ On the
other hand,
$$\hat \Theta(g\circ
f)=\int_{\M}\left|\varphi-\chi\circ g\circ f \right|d\mu=4\pi.$$
Hence, the inequality $\hat \Theta(g\circ f)\le \hat
\Theta(f)+\hat \Theta(g)$ does not hold. This fact prevents the
function $\inf_{f\in H\left({\M},{\N}\right)} \hat \Theta(f)$ from
being a pseudo-distance, since we do not get the triangular
inequality. Furthermore this function is not symmetric for all
couples of size pairs.

Other dissimilarity measures
based on some integral of the change of the measuring functions are object of a on-going research. Preliminary results on this subject can be found in \cite{FrLa07}.}
\end{rem}

For more details about natural
pseudo-distances between size pairs, the reader is referred to \cite{Donatini}, 
\cite{donatini2004} and \cite{donatini2007}.

Before stating the main result of this section (Th.~\ref{nsd}), in
Th.~\ref{lem:minmax} we show that the $\inf$ and the $\sup$ in
the definition of matching distance are actually attained, that is
to say, a matching $\sigma$ exists for which $d_{match}(\ell^*_1,\ell^*_2)=\min_\sigma\max_i
d(a_i,b_{\sigma(i)})$. Every such  matching will henceforth be called {\em optimal}.

\begin{thm}{\rm (Optimal Matching Theorem)}\label{lem:minmax}
Let $(a_i)$ and $(b_i)$ be two representative sequences of points for the reduced size functions $\ell^*_1$ and $\ell^*_2$ respectively. Then
the matching distance
between $\ell^*_1$ and $\ell^*_2$ is equal to the number
$\min_\sigma\max_i d(a_i,b_{\sigma(i)})$,
where $i$ varies in $\bn$ and $\sigma$ varies among all the bijections from $\bn$ to $\bn$.
\end{thm}

\begin{pf}
By the definition of the pseudo-distance $d$, we can confine ourselves to considering  only the bijections $\sigma$, such that $\sigma(0)=0$. In this way, by Prop.~\ref{corn=pos} (Localization of cornerpoints),  we obtain that
$\sup_{i\in\bn} d(a_i,b_{\sigma(i)})< +\infty$.

Let us first see that $\sup_{i\in\bn} d(a_i,b_{\sigma(i)})=\max_{i\in\bn}
d(a_i,b_{\sigma(i)})$, for any such bijection $\sigma$. This  is true because proper cornerpoints belong to a bounded set,
 and the accumulation
points for the  set of  cornerpoints of a reduced size function (if
any) cannot belong to $\Delta^+$, but only to  $\Delta$ (see
Prop.~\ref{corncount}, Local finiteness of cornerpoints). By definition, for any $p$ and $p'$ in
$\Delta$, $d(p,p')=0$, and hence the claim is proved.

Let us now prove that $\inf_\sigma\max_i
d(a_i,b_{\sigma(i)})=\min_\sigma\max_i d(a_i,b_{\sigma(i)})$. Let
us set $s:=\inf_\sigma\max_i d(a_i,b_{\sigma(i)})$.  According to
Prop.~\ref{corncount}, if $s=0$, then the cornerpoints of
$\ell^*_1$ coincide with those of $\ell^*_2$, and their
multiplicities are the same, implying that  the claim is true. Let
us consider the case when $s>0$. Let $J_1:=\{i\in
\bn:d(a_i,\Delta)>s\}$ and $J_2:=\{i\in \bn:d(a_i,\Delta)\le s\}$ (see Fig.~\ref{Fig:js}).
By Prop.~\ref{corncount}, $J_1$ contains only a finite number of
elements,  and therefore there exists a
real positive number $\epsilon$ for which
$d(a_i,\Delta)>s+\epsilon$ for each $i\in J_1$.

\par
\begin{figure}
\psfrag{d}{$d=s$}
\psfrag{J}{$a(J_1)$}
\psfrag{H}{$a(J_2)$}
\psfrag{s}{$\bar\sigma$}
\psfrag{D}{$\Delta$}
\centerline{\hbox{
 \includegraphics[width=2.5in]{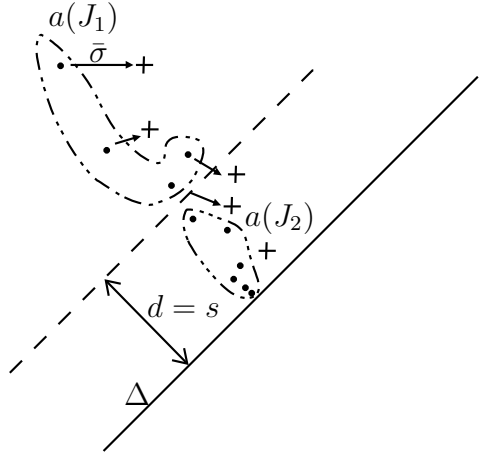}
}} \caption{The sets $a(J_1)$ and $a(J_2)$ of cornerpoints corresponding to the sets $J_1$ and $J_2$ used in the proof of
Th.~\ref{lem:minmax}. The sequence $(a_i)$ is represented by dots and the sequence $(b_i)$ by crosses.}\label{Fig:js}
\end{figure}
\par

Let us consider the set $\Sigma$ of all the
injective functions $\sigma:J_1\rightarrow \bn$ such that
$\max_{i\in J_1}d(a_i,b_{\sigma(i)})<s+\epsilon/2$. This set is
non-empty by the definition of $s$, and contains only a finite number
of injections because  $J_1$ is finite,  and for each $i\in J_1$
the set $\{ j\in \bn: d(a_i,b_j)<s+\epsilon/2\}$ is
 finite. Thus we can take an injection ${\bar \sigma}:J_1\rightarrow \bn$ that realizes the minimum of $\max_{i\in J_1}d(a_i,b_{\sigma(i)})$ as $\sigma$ varies in $\Sigma$.
Obviously, $\max_{i\in J_1}d(a_i,b_{{\bar \sigma}(i)})\le s$, by the definition of $s$.

Moreover, we can take an injection $\hat \sigma:J_2\rightarrow \bn$
such that $\max_{i\in J_2}d(a_i,b_{\hat\sigma(i)})\le s$, because,
for every $i\in J_2$, we can choose a different index $j$, such
that $b_j\in\Delta$.
Since $\bn=J_1\cup J_2$ and $Im(\bar \sigma)\cap Im(\hat\sigma)=\emptyset$,  we can construct an injection $f:\bn\rightarrow \bn$ such that
 each displacement between $a_i$ and $b_{f(i)}$ is not greater
 than $s$, by setting $f(i)=\bar\sigma(i)$ for $i\in J_1$, and $f(i)=\hat \sigma(i)$ for $i\in J_2$.
Analogously, we can construct an injection $g:\bn\rightarrow \bn$
such that each displacement between $b_i$ and  $a_{g(i)}$ is not
greater than  $s$. Therefore, by the Cantor-Bernstein Theorem, there is a
bijection $l:\bn\rightarrow \bn$ such that $d(a_i,b_{l(i)})\le s$
for every index $i$, and so the theorem is proved.
\qed\end{pf}

\begin{thm}\label{nsd}
Let $\epsilon \ge 0$ be a real number and let $(\M,\p)$ and $(\N,\s)$ be two size pairs with $\M$ and $\N$ homeomorphic.
Then
$$d_{match}(\rsfp, \rsfn)\le\inf_{h}\max _{P\in \M}|\p(P)-\s(h(P))|,$$
where $h$ ranges among all the homeomorphisms from $\M$ to $\N$.
\end{thm}

\begin{pf}
We begin by observing that $\rsfn= \ell^*_{(\M,\s\circ h)}$, where $h:\M\rightarrow \N$ is any
 homeomorphism between $\M$ and $\N$. Moreover, for each homeomorphism $h$, by applying Th.~\ref{main} with $\epsilon= \max _{P\in \M}|\p(P)-\s(h(P))|$,
we have
$$d_{match}(\rsfp,\ell^*_{(\M,\s\circ h)})\le \max _{P\in \M}|\p(P)-\s(h(P))|.$$
 Since this is true for any homeomorphism $h$ between $\M$ and $\N$,
it  immediately follows that $d_{match}(\rsfp, \rsfn)\le\inf_{h}\max _{P\in \M}|\p(P)-\s(h(P))|$.
\qed\end{pf}

\section{Construction of size pairs with given reduced size functions and natural pseudo-distance}

The following lemma states that it is always possible to construct two size pairs such that their reduced  size functions
are assigned in advance, and their natural pseudo-distance equals the matching distance between the corresponding reduced size functions.

This result evidently allows us to  deduce that the lower bound for the natural pseudo-distance given by the matching distance is sharp. Furthermore, in the next section, we will exploit this lemma to conclude that, although other distances between reduced size functions
could in principle be thought of in order to obtain a lower bound for the natural pseudo-distance, in practice it would be  useless since the matching distance furnishes the best estimate.

The proof of this lemma is rather technical, so we anticipate the underlying idea.  We aim at constructing a rectangle $R$ and two measuring functions $\tilde\p$  and $\tilde \s$ on $R$, such that $\ell^*_{(R,\tilde \p)}= \rsfp$ and $\ell^*_{(R,\tilde \s)}= \rsfn$ for given $(\M,\p)$ and $(\N,\s)$. To this aim, we fix an optimal matching between cornerpoints of $\rsfp$ and $\rsfn$ that achieves $d_{match}( \rsfp,\rsfn)$.
We begin by defining $\tilde\p$ and $\tilde\s$ as linear functions on $R$, with minima respectively at height $\min\p$ and $\min\s$. Thus, the corresponding reduced size functions have cornerpoints at infinity coinciding with those of $\rsfp$ and $\rsfn$  respectively, and no proper cornerpoints.   Next, for each  proper cornerpoint $(x,y)$ for $\rsfp$, considered with its multiplicity,  we modify the graph of $\tilde\p$  by digging a pit with bottom  at height $x$ and top at height $y$ (see Fig.~\ref{total}). This creates a proper cornerpoint for  $\ell^*_{(R,\tilde \p)}$ at $(x,y)$. Of course, we take care that  this occurs at different points of $R$ for different cornerpoints of $\ell^*_{(R,\tilde \p)}$.  Now we modify the graph of $\tilde\s$  exactly at the same point of $R$ as   for $\tilde\p$, as follows. If the  cornerpoint $(x,y)$  for $\rsfp$ is matched to a cornerpoint $(x',y')$ of $\rsfn$, we dig a pit in the graph of $\tilde\s$ with bottom  at height $x'$ and top at height $y'$. This creates a proper cornerpoint for  $\ell^*_{(R,\tilde \s)}$ at $(x',y')$.
 Otherwise, if the  cornerpoint $(x,y)$  for $\rsfp$ is matched to a point of the diagonal,  we construct a plateau in the graph of $\tilde \s$ at height $\frac{x+y}{2}$. This does not introduce new cornerpoints for $\ell^*_{(R,\tilde \s)}$. We repeat the procedure backwards, for cornerpoints of $\rsfn$ that are matched to points of the diagonal in $\rsfp$. Although this is the idea underlying the construction of the measuring functions, in our proof of Lemma~\ref{cubi} we shall confine ourselves to giving only their  final definition.

Now, in order to show that $d_{match}(\ell^*_{(R,\tilde \p)}, \ell^*_{(R,\tilde \s)})$ is equal to the natural pseudo-distance between  $(R,\tilde\p)$ and $(R,\tilde\s)$, we consider the identity function on $R$. We measure the difference between  $\tilde\p$ and $\tilde\s$ at each point of $R$. By construction, this difference is greater at pits and  plateaux than elsewhere, that is to say, corresponds to the matching of cornerpoints.  Hence, the identity homeomorphism of $R$ achieves a change in the measuring functions $\tilde \p$ and $\tilde \s$ equal to the matching distance between $\ell^*_{(R,\tilde\p)}$ and  $\ell^*_{(R,\tilde\s)}$. This implies that the matching distance is equal to the natural pseudo-distance between  $(R,\tilde\p)$ and $(R,\tilde\s)$, and the identity function on $R$ is exactly the homeomorphism attaining the natural pseudo-distance.

The last step of the proof is to enlarge the rectangle $R$ to a topological $2$-sphere  and to extend the measuring functions to this surface, without modifying the corresponding reduced size functions and the natural pseudo-distance.

\begin{lem}\label{cubi}
Let $\rsfp$ and $\rsfn$ be two reduced size functions. There always exist two size pairs $(\M',\p')$ and $(\M',\s')$, with $\M'$ homeomorphic to a $2$-sphere, such that the following statements hold:
\begin{enumerate}
\item $\ell^*_{(\M',\p')}= \rsfp$;
\item $\ell^*_{(\M',\s')}= \rsfn$;
\item $d_{match}(\ell^*_{(\M',\p')}, \ell^*_{(\M',\s')})=\inf_h\max_{P\in \M'}|\p'(P)-\s'(h(P))|$,  $h$ sweeping the set
$H(\M',\M')$ of all self-homeomorphisms of $\M'$, that is to say the matching distance equals the natural pseudo-distance;
\item $\inf_h\max_{P\in \M'}|\p'(P)-\s'(h(P)|=\max_{P\in \M'}|\p'(P)-\s'(P)|$, that is to say, the identity homeomorphism attains the natural pseudo-distance.
\end{enumerate}
\end{lem}
\medskip

\begin{pf}
 Possibly by swapping the size pairs, we can assume that $\min\p\le\min\s$.

Let $(a_i)$ and $(b_i)$ be two representative sequences for  $\rsfp$ and $\rsfn$, respectively. It is not restrictive to assume that the identical bijection $i\mapsto i$ is an optimal matching such that
$d_{match}(\ell^*_{(\M,\p)}, \ell^*_{(\N,\s)})$ is attained.  We define $$I:=\{i\in\bn-\{0\}: a_i\notin\Delta\}, \quad J:=\{j\in\bn-\{0\}: b_j\notin\Delta\}.$$

For every $i>0$, let  $a_i=(x_i,y_i)$ and $b_i=(x'_i,y'_i)$. Then
$$
\lim_{i\to\infty}(y_i-x_i)=0, \  \ \ \ \ \ \ \ \ \  \lim_{i\to\infty}(y'_i-x'_i)=0,
 $$
because of Prop.~\ref{corncount} (Local finiteness of cornerpoints).

We choose a real number $S$ with $S>\max\{\max\p,\max\s\}$. Necessarily $S>\min\p$.
Moreover, we define $(\epsilon_i)_{i\in\bn-\{0\}}$ to be a sequence of positive real number tending to $0$ and such  that, for every $i\in\bn-\{0\}$, it holds that:
$$\left[\frac{x_i+y_i}{2}-\epsilon_i,\frac{x_i+y_i}{2}+\epsilon_i\right]\subseteq (\min\p,S).$$
 Finally, let $R:=[0,1]\times [\min\p,S]\subseteq\br^2$.

We define $\tilde\p:R\rightarrow \br$ by  the following conditions:

 (i) Let $i\in I$ and  $a_i=(x_i,y_i)$. Then $\tilde\p(\frac{1}{3i},y)$ is the piecewise linear function in the variable $y\in [\min\p,S]$ whose graph, consisting of  four segments, is represented in Fig.~\ref{primo}(left).
Furthermore, $\tilde\p(\frac{1}{3i-1},y)$ and $\tilde\p(\frac{1}{3i+1},y)$ are the piecewise linear functions in the variable $y\in [\min\p,S]$, whose graph (the same) consists of  three segments, as represented in Fig.~\ref{primo}(right).
\medskip

\begin {figure}
  \psfrag{A}[l][l][1][90]{$\min\p$}
  \psfrag{E}{$\min\p$}
  \psfrag{B}[l][l][1][90]{$\frac{x_i+y_i}{2}$}
  \psfrag{C}[l][l][1][90]{$\frac{x_i+y_i}{2}-\epsilon_i$}
  \psfrag{D}[l][l][1][90]{$\frac{x_i+y_i}{2}+\epsilon_i$}
  \psfrag{S}{$S$}
  \psfrag{X}{$x_i$}
  \psfrag{Y}{$y_i$}
 \psfrag{L}{$y$}
  \psfrag{F}{$\tilde\p(\frac{1}{3i},y)$}
\psfrag{G}{$\tilde\p(\frac{1}{3i\pm 1},y)$}
  \begin{center}\begin{tabular}{cc}
\includegraphics[height=5cm]{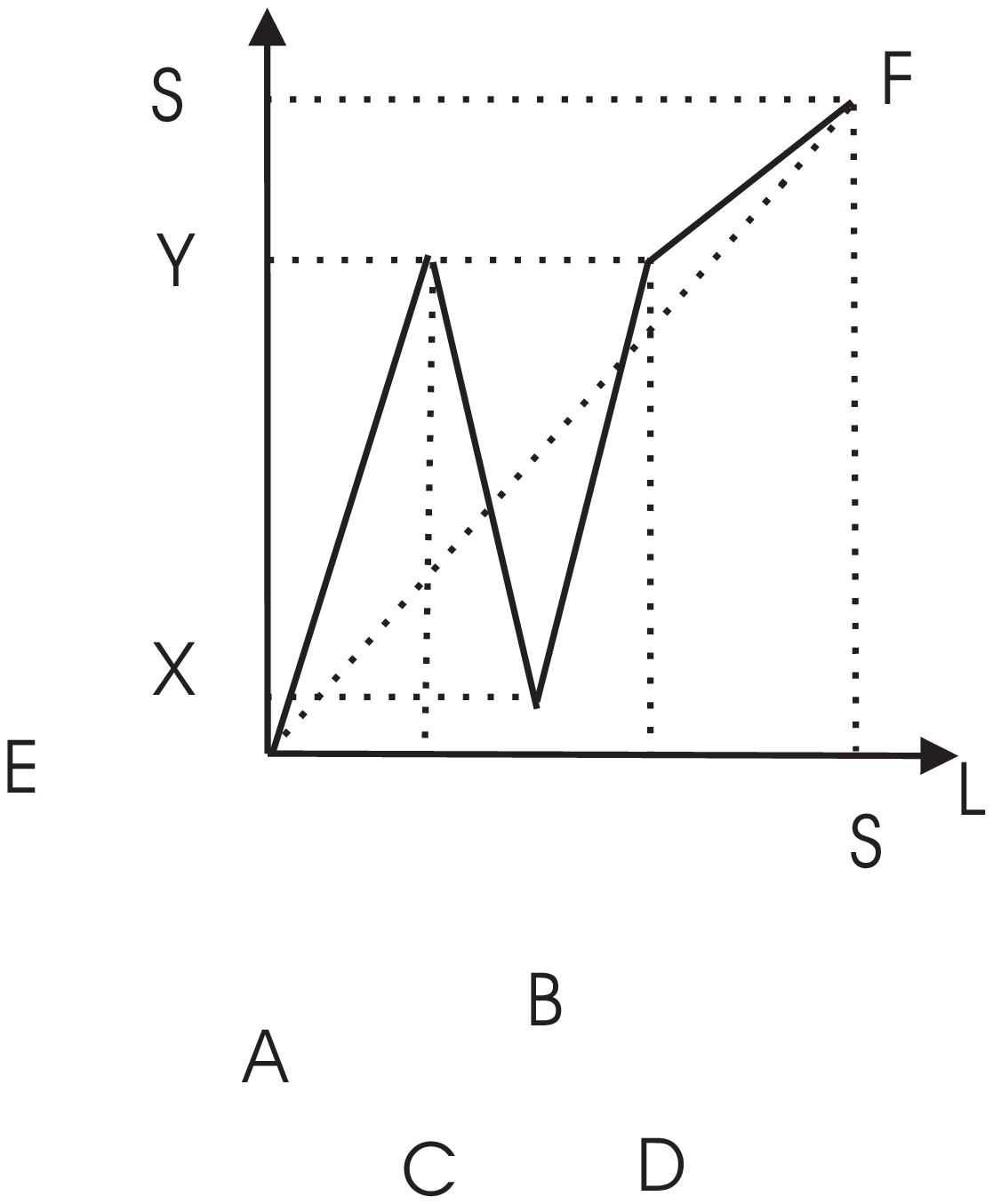} \quad &
\quad \includegraphics[height=5cm, bb=120 442 386 816]{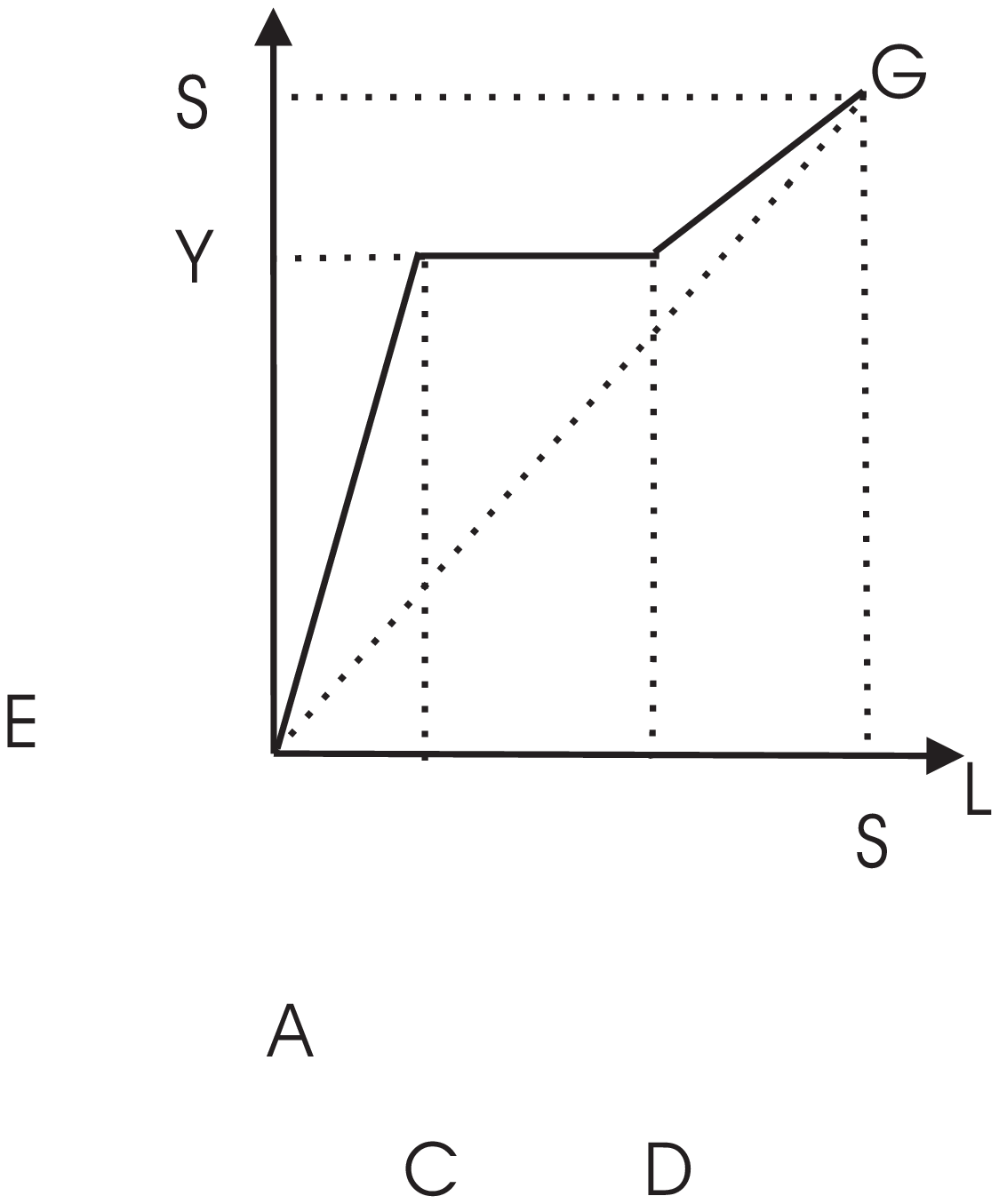}
\end{tabular} \end{center}
\caption {Case $i\in I$. Left: The graph of the function $\tilde\p(\frac{1}{3i},y)$. Right: The graph of the functions $\tilde\p(\frac{1}{3i\pm 1},y)$.}
  \label{primo}
\end {figure}

 (ii) Let $i\in J-I$ and $b_i=(x'_i,y'_i)$. Then $\tilde\p(\frac{1}{3i},y)$ and  $\tilde\p(\frac{1}{3i\pm 1},y)$ are  the piecewise linear functions in the variable $y\in [\min\p,S]$ whose graph, consisting of  three segments, is represented in Fig.~\ref{three2}. Let us observe that $\frac{ x'_i+y'_i}{2}>\min\p$, since we are assuming $\min\p\le \min\s$.
\medskip

\begin {figure}
  \psfrag{A}[l][l][1][90]{$\min\p$}
  \psfrag{E}{$\min\p$}
  \psfrag{B}{$\frac{x'_i+y'_i}{2}$}
  \psfrag{C}[l][l][1][90]{$\frac{x'_i+y'_i}{2}-\epsilon_i$}
  \psfrag{D}[l][l][1][90]{$\frac{x'_i+y'_i}{2}+\epsilon_i$}
  \psfrag{S}{$S$}
 \psfrag{L}{$y$}
  \psfrag{F}{$\tilde\p(\frac{1}{3i},y)$, $\tilde\p(\frac{1}{3i\pm 1},y)$}
 \begin{center}
\includegraphics[height=5cm]{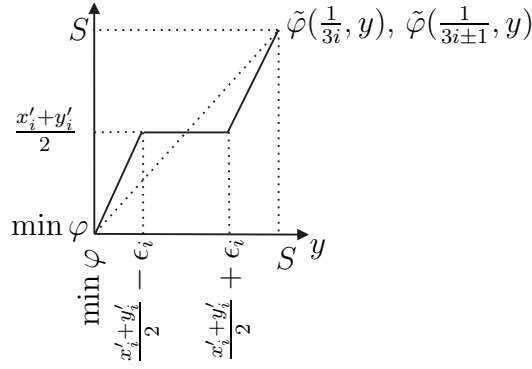}
\caption {Case $i\in J-I$. The graph of the functions $\tilde\p(\frac{1}{3i},y)$ and $\tilde\p(\frac{1}{3i\pm 1},y)$.}
  \label{three2}
\end{center}
\end {figure}

 (iii) $\tilde\p(0,y)$ and $\tilde\p(1,y)$,  as functions in the variable $y\in [\min\p,S]$, are defined as the identity function $y\mapsto y$.
\medskip

(iv) For any $x\in [0,1]$ where not already defined, $\tilde\p(x,y)$ is defined by linear
extension with respect to the variable $x$.
\medskip

The  aspect of the graph of $\tilde \p$ in correspondence of a cornerpoint is represented in Fig.~\ref{total}.

\begin {figure}
 \psfrag{x}{$x\in[0,1]$}
 \psfrag{y}{$y\in[\min\p,S]$}
 \begin{center}
\includegraphics[height=7.5cm]{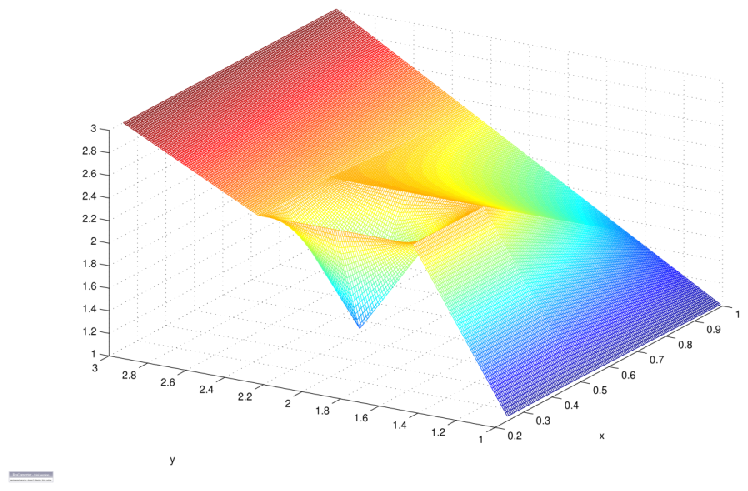}
\caption {A ``pit'' of the function $\tilde\p:R\rightarrow \br$.}
  \label{total}
\end{center}
\end {figure}


Let us observe that, by construction, $\tilde\p(x,y)$ is continuous everywhere in $R$, except possibly at $x=0 $. Actually, $\tilde\p(x,y)$ is continuous also at $x=0 $. Indeed, when $i$ tends to infinity,  $\epsilon_i\to 0$ and
 $(y_i-x_i)\to 0$. Thus the functions $\tilde\p(\frac{1}{i},y)$ with $i>0 $ tend to the identity function in the $\sup$ norm.

It is not difficult to check that $\ell^*_{(R,\tilde\p)}=\rsfp$.

Let us now analogously define the function $\tilde\s:R\rightarrow \br$, where $R$ is still defined as $[0,1]\times [\min\p,S]\subseteq\br^2$, by means of the following conditions:

(i')  Let $i\in J$ and $b_i=(x'_i,y'_i)$. Then $\tilde\s(\frac{1}{3i},y)$ is the piecewise linear function in the variable $y\in [\min\p,S]$ whose graph, constituted of  four segments, is represented in Fig.~\ref{secondo}(left).
Furthermore, $\tilde\s(\frac{1}{3i-1},y)$ and $\tilde\s(\frac{1}{3i+1},y)$ are the piecewise linear functions in the variable $y\in [\min\p,S]$ whose graph (the same), constituted of  three segments, is represented in Fig.~\ref{secondo}(right).
\medskip

\begin {figure}
  \psfrag{A}[l][l][1][90]{$\min\p$}
  \psfrag{B}[l][l][1][90]{$\frac{x_i+y_i}{2}$}
  \psfrag{C}[l][l][1][90]{$\frac{x_i+y_i}{2}-\epsilon_i$}
  \psfrag{D}[l][l][1][90]{$\frac{x_i+y_i}{2}+\epsilon_i$}
  \psfrag{S}{$S$}
  \psfrag{X}{$x'_i$}
  \psfrag{Y}{$y'_i$}
 \psfrag{L}{$y$}
\psfrag{E}{$\min\s$}
  \psfrag{F}{$\tilde\s(\frac{1}{3i},y)$}
 \psfrag{G}{$\tilde\s(\frac{1}{3i\pm 1},y)$}
 \begin{center}\begin{tabular}{cc}
 \includegraphics[height=5cm]{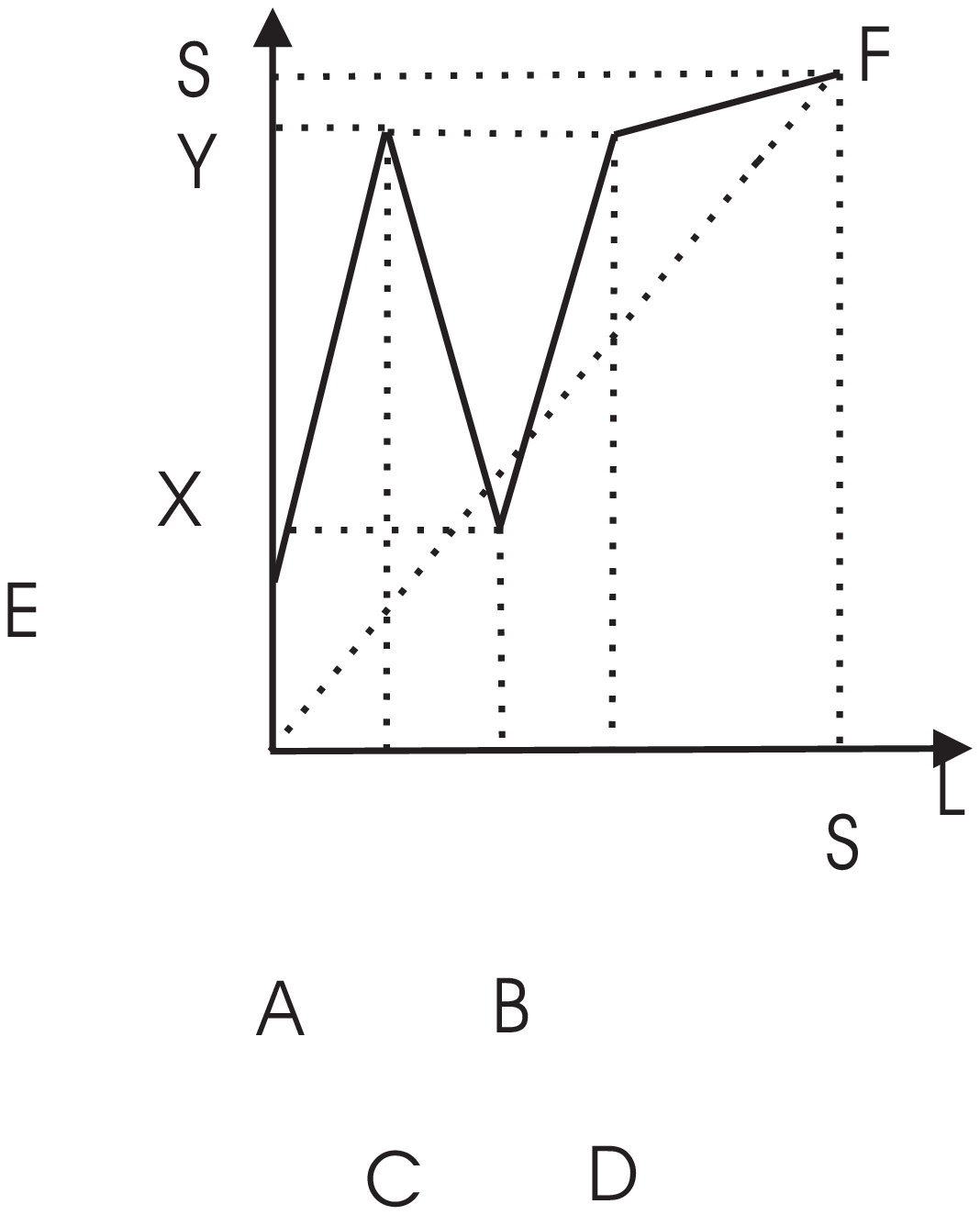}\quad &
\quad \includegraphics[height=5cm]{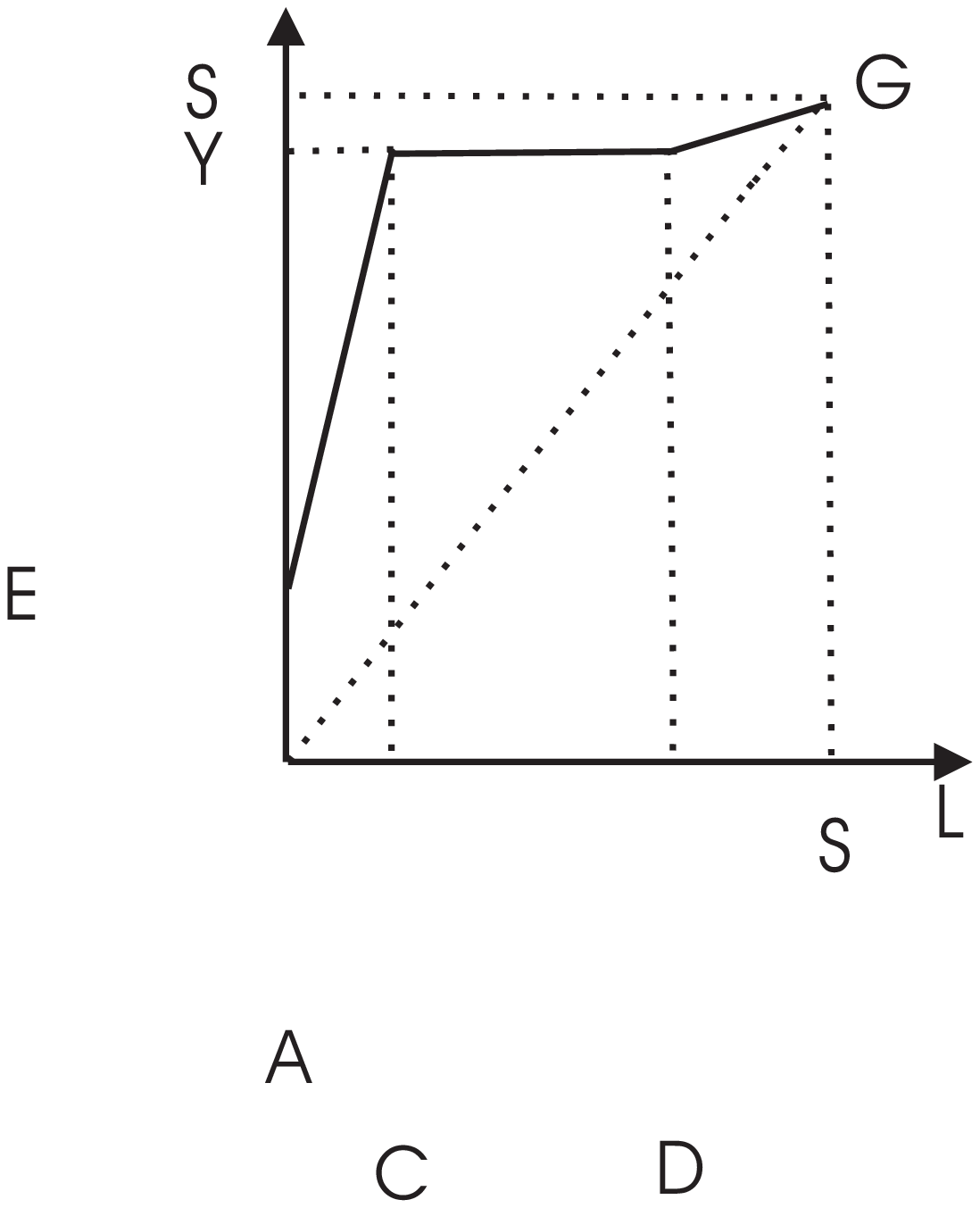}\end{tabular}\end{center}
\caption {Case $i\in J$. Left: The graph of the function $\tilde\s(\frac{1}{3i},y)$.  Right: The graph of the functions $\tilde\s(\frac{1}{3i\pm 1},y)$.}
  \label{secondo}
\end {figure}

(ii') Let $i\in I-J$ and  $a_i=(x_i,y_i)$.  Then, if $\frac{x_i+y_i}{2}>\min\s$, we take  $\tilde\s(\frac{1}{3i},y)$ and
$\tilde\s(\frac{1}{3i\pm 1},y)$ to be the piecewise linear functions in the variable $y\in [\min\p,S]$, whose graph, constituted of  three segments, is represented in Fig.~\ref{terzo}(left). Otherwise, if $\frac{x_i+y_i}{2}\le\min\s$, we take  $\tilde\s(\frac{1}{3i},y)$ and $\tilde\s(\frac{1}{3i\pm 1},y)$ to be the piecewise linear functions in the variable $y\in [\min\p,S]$ whose graph, constituted of  two segments, is represented in Fig.~\ref{terzo}(right),
\medskip

\begin {figure}
  \psfrag{A}[l][l][1][90]{$\min\p$}
  \psfrag{B}[l][l][1][90]{$\frac{x_i+y_i}{2}$}
  \psfrag{C}[l][l][1][90]{$\frac{x_i+y_i}{2}-\epsilon_i$}
  \psfrag{D}[l][l][1][90]{$\frac{x_i+y_i}{2}+\epsilon_i$}
\psfrag{H}{$\frac{x_i+y_i}{2}$}
  \psfrag{S}{$S$}
 \psfrag{L}{$y$}
\psfrag{E}{$\min\s$}
  \psfrag{F}{$\tilde\s(\frac{1}{3i},y)$}
\psfrag{G}{$\tilde\s(\frac{1}{3i\pm 1},y)$}
 \begin{center}\begin{tabular}{cc}
\includegraphics[height=5cm]{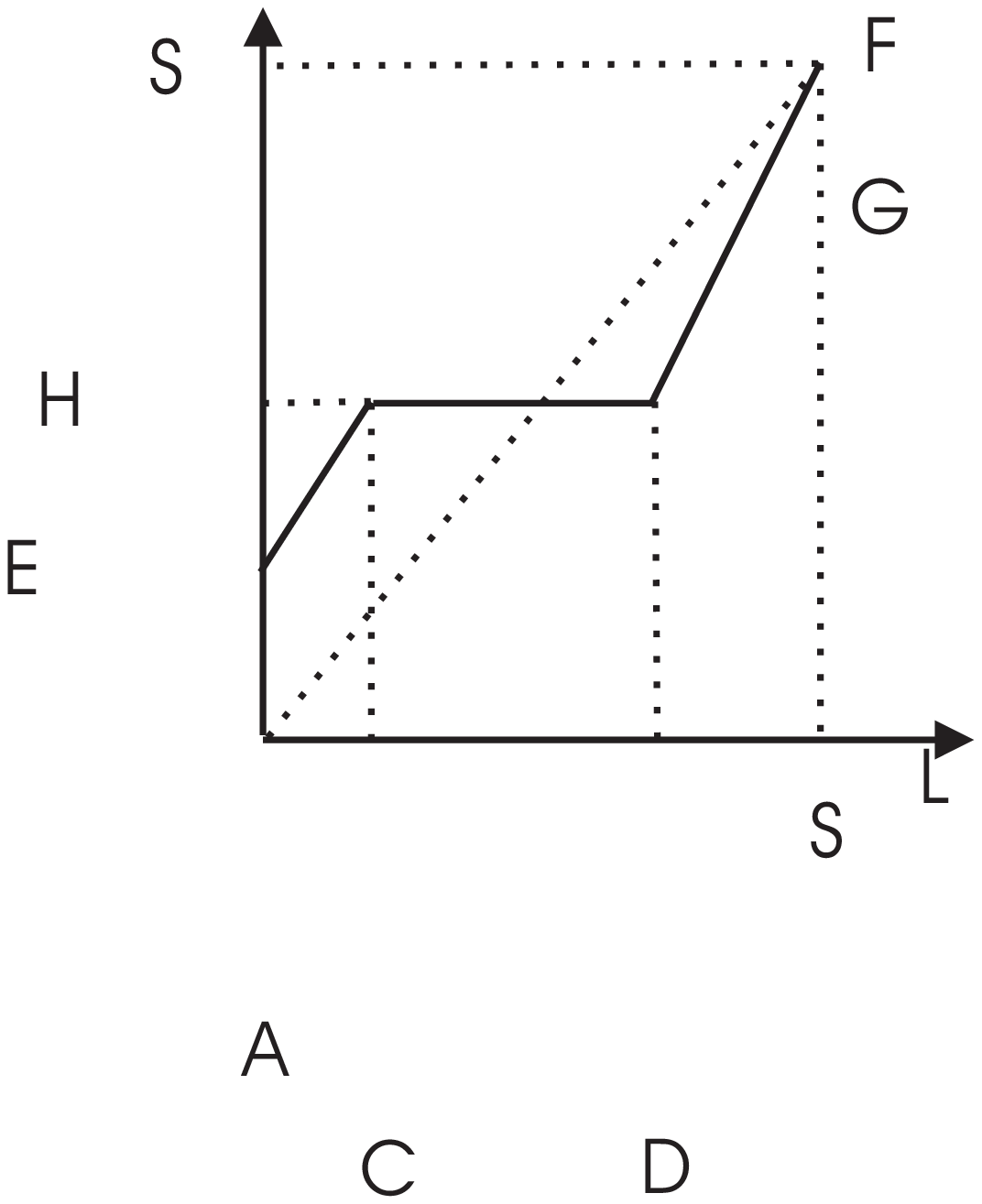} \quad \quad&
\quad \quad \includegraphics[height=5cm]{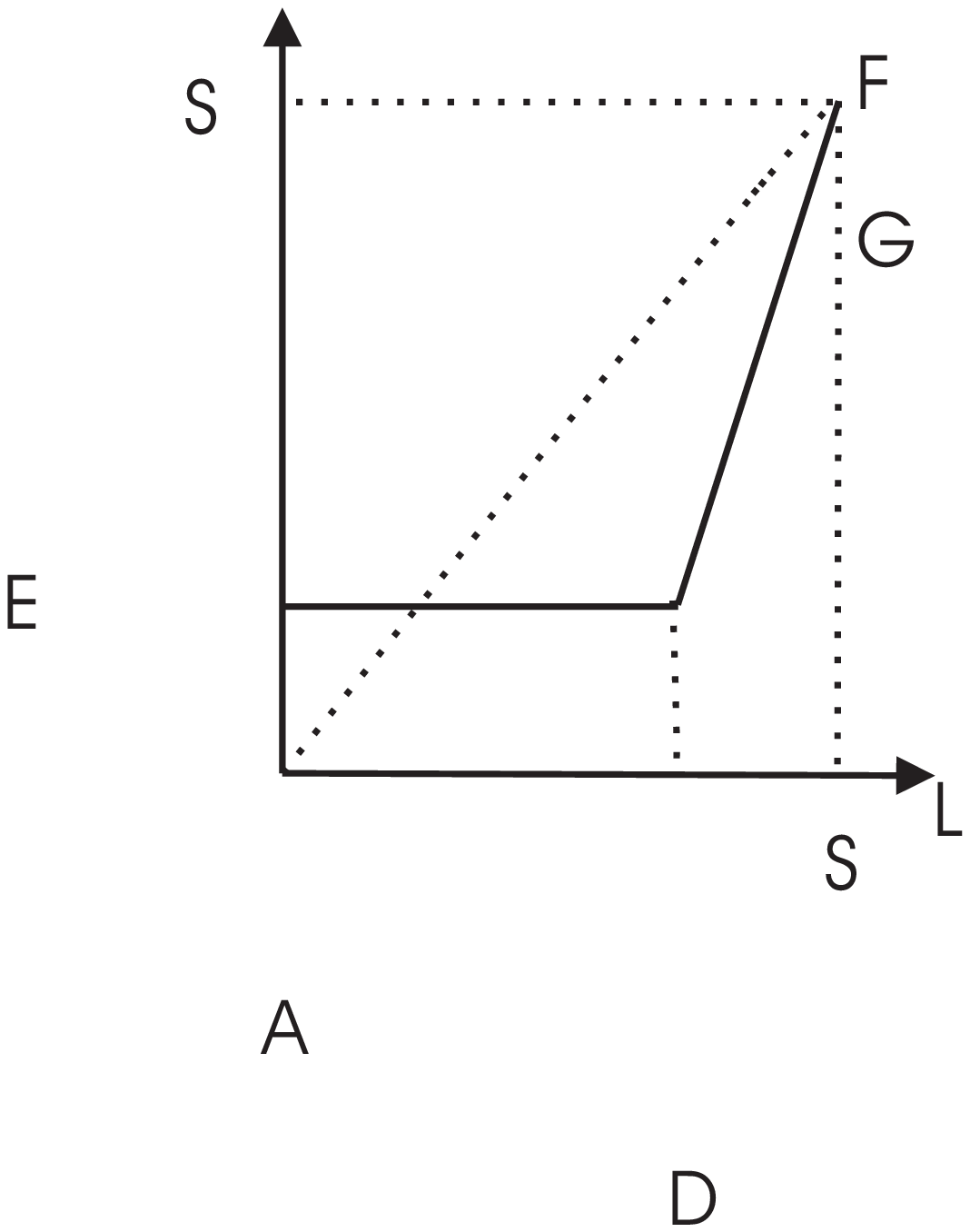}\end{tabular}\end{center}
\caption {Case $i\in I-J$. Left: The graph of the functions $\tilde\s(\frac{1}{3i},y)$ and $\tilde\s(\frac{1}{3i\pm 1},y)$, when  $\frac{x_i+y_i}{2}>\min\s$. Right: The graph of the functions $\tilde\s(\frac{1}{3i},y)$ and $\tilde\s(\frac{1}{3i\pm 1},y)$, when $\frac{x_i+y_i}{2}\le\min\s$.}
  \label{terzo}
\end {figure}

(iii') $\tilde\s(0,y)$ and $\tilde\s(1,y)$,  as functions in the variable $y\in [\min\p,S]$, are both defined as the  function $y\mapsto \frac{S-\min\s}{S-\min\p}(y-\min\p)+\min\s$, that is to say, the linear function that takes value $\min\s$, when $y=\min\p$, and value $S$, when $y=S$.

(iv') For any $x\in [0,1]$, where not already defined, $\tilde\s(x,y)$ is defined by linear
extension with respect to the variable $x$.

Again, one sees that  $\tilde\s:R\rightarrow \br$ is continuous and $\ell^*_{(R,\tilde\s)}=\rsfn$.
\medskip

Let us now show that $d_{match}(\ell^*_{(R,\tilde \p)}, \ell^*_{(R,\tilde\s)})=\max_{P\in R}|\tilde\p(P)-\tilde\s(P)|$.
It may be useful to recall that $d_{match}(\ell^*_{(R,\tilde \p)}, \ell^*_{(R,\tilde\s)})=\max_{i\in \bn}d(a_i,b_i)$,  where, if $a_i=(x_i,y_i)$ and $b_i=(x'_i,y'_i)$,

\begin{tabular}{ll}
if $i=0$ & $d(a_i,b_i)=\min\s-\min\p$;\\
if $i\in I\cap J$, & $d(a_i,b_i)=\max\{|y'_i-y_i|,|x'_i-x_i|\}$;\\
if $i\in I- J$, & $d(a_i,b_i)=\frac{y_i-x_i}{2}$;\\
if $i\in J-I$, & $d(a_i,b_i)=\frac{y'_i-x'_i}{2}$;\\
if $i\notin I\cup J\cup \{0\}$, & $d(a_i,b_i)=0$.
\end{tabular}

As for $\max_{P\in R}|\tilde\p(P)-\tilde\s(P)|$, notice that the maximum must be attained at a point $(x,y)\in R$,
where either $x=\frac{1}{3i}$ or $x=\frac{1}{3i\pm 1}$, with $i>0$, or $x=0$, or $x=1$. Indeed, $\tilde\p$ and $\tilde\s$ are linearly defined in the variable $x$ elsewhere. Let us  consider  the various instances separately.

For $x=0$ and $x=1$, it holds that
$$\max_{y\in [\min\p,S]}|\tilde\p(x,y)-\tilde\s(x,y)|= \min\s-\min\p=d(a_0,b_0).$$

For $x=\frac{1}{3i}$ or $x=\frac{1}{3i\pm 1}$, different cases are possible.

\underline{Case  $ \mathit{i\in I\cap J}$}. By looking at the graphs of $\tilde\p(\frac{1}{3i},y)$ when $i\in I$ and $\tilde\s(\frac{1}{3i},y)$ when $i\in J$, we immediately see that
\begin{eqnarray*}
\lefteqn{\max_{y\in [\min\p,S]}\left|\tilde\p\left(\frac{1}{3i},y\right)-\tilde\s\left(\frac{1}{3i},y\right)\right|=}\\
& & =\max \{|y'_i-y_i|,|x'_i-x_i|, \min\s-\min\p\}=\\
& & = \max\{d(a_i,b_i),d(a_0,b_0)\}.
\end{eqnarray*}
By looking at the graphs of $\tilde\p(\frac{1}{3i\pm 1},y)$ when $i\in I$ and $\tilde\s(\frac{1}{3i\pm 1},y)$ when $i\in J$, we see that
\begin{eqnarray*}
\lefteqn{\max_{y\in [\min\p,S]}\left|\tilde\p\left(\frac{1}{3i\pm 1},y\right)-\tilde\s\left(\frac{1}{3i\pm 1},y\right)\right|=}\\
& & =\max \{|y'_i-y_i|, \min\s-\min\p\}\le \\
& & \le  \max\{d(a_i,b_i),d(a_0,b_0)\}.
\end{eqnarray*}

\underline{ Case $\mathit{i\in I- J}$.} If $\frac{x_i+y_i}{2}>\min\s$, then
\begin{eqnarray*}
\lefteqn{\max_{y\in [\min\p,S]}\left|\tilde\p\left(\frac{1}{3i},y\right)-\tilde\s\left(\frac{1}{3i},y\right)\right|=}\\
& & =\max \left\{\left|y_i-\frac{x_i+y_i}{2}\right|,\left|x_i-\frac{x_i+y_i}{2}\right|, \min\s-\min\p\right\}=\\
& & =\max \left\{\left|\frac{y_i-x_i}{2}\right|, \min\s-\min\p\right\}=\\
& & = \max\{d(a_i,b_i),d(a_0,b_0)\}.
\end{eqnarray*}
Or else, if $\frac{x_i+y_i}{2}\le \min\s$, then
$$|y_i-\min\s|< |x_i-\min\s|,$$
 because $x_i< y_i\le 2\min\s-x_i$ and hence
$x_i-\min\s<y_i-\min\s\le\min\s-x_i.$
Moreover,  $|x_i-\min\s|=\min\s-x_i\le \min\s-\min\p$.
Therefore,
\begin{eqnarray*}
\lefteqn{\max_{y\in [\min\p,S]}\left|\tilde\p\left(\frac{1}{3i},y\right)-\tilde\s\left(\frac{1}{3i},y\right)\right|=}\\
& & =\max \{|y_i-\min\s|, |x_i-\min\s|,\min\s-\min\p\}= \\
& & =\min\s-\min\p= d(a_0,b_0).
\end{eqnarray*}
Analogously, if $\frac{x_i+y_i}{2}>\min\s$, then
$$\max_{y\in [\min\p,S]}\left|\tilde\p\left(\frac{1}{3i\pm 1},y\right)-\tilde\s\left(\frac{1}{3i\pm 1},y\right)\right|= \max\{d(a_i,b_i),d(a_0,b_0)\},$$
or, if $\frac{x_i+y_i}{2}\le \min\s$, then
$$\max_{y\in [\min\p,S]}\left|\tilde\p\left(\frac{1}{3i\pm 1},y\right)-\tilde\s\left(\frac{1}{3i\pm 1},y\right)\right|= d(a_0,b_0).$$

\underline{ Case $\mathit{i\in J-I}$.} We have that
\begin{eqnarray*}
\lefteqn{\max_{y\in [\min\p,S]}\left|\tilde\p\left(\frac{1}{3i},y\right)-\tilde\s\left(\frac{1}{3i},y\right)\right|=}\\
& & =\max \left\{\left|y'_i-\frac{x'_i+y'_i}{2}\right|,\left|x'_i-\frac{x'_i+y'_i}{2}\right|, \min\s-\min\p\right\}=\\
& & =\max \left\{\left|\frac{y'_i-x'_i}{2}\right|, \min\s-\min\p\right\}=\\
& & = \max\{d(a_i,b_i),d(a_0,b_0)\},
\end{eqnarray*}
and
\begin{eqnarray*}
\max_{y\in [\min\p,S]}\left|\tilde\p\left(\frac{1}{3i\pm 1},y\right)-\tilde\s\left(\frac{1}{3i\pm 1},y\right)\right|= \max\{d(a_i,b_i),d(a_0,b_0)\}.
\end{eqnarray*}

From all these facts we deduce that
$$\max_{P\in R}|\tilde\p(P)-\tilde\s(P)|= d_{match}(\ell^*_{(R,\tilde\p)},\ell^*_{(R,\tilde\s)}).$$

To complete the proof, it is now sufficient to extend the above arguments to a topological $2$-sphere
.
To do this, let us take
$$\M':=\partial(R\times [0,1])=\partial([0,1]\times[\min\p,S]\times [0,1]).$$
Obviously, $\M'$ is homeomorphic to a $2$-sphere.

 Moreover, by identifying  $R\times \{0\}$ with $R$, let us define $\p':\M'\rightarrow\br$ as follows:
$\p'_{|R\times \{0\}}\equiv\tilde\p$, $\p'_{|[0,1]\times\{\min\p\}\times [0,1]}\equiv\min\p$, $\p'_{|[0,1]\times\{S\}\times [0,1]}\equiv S$, and  let us define $\p'$ by linear extension in $y\in[\min\p,S]$ elsewhere. Finally, let us define  $\s':\M'\rightarrow\br$ analogously: $\s'_{|R\times \{0\}}\equiv\tilde\s$, $\s'_{|[0,1]\times\{\min\p\}\times [0,1]}\equiv\min\s$, $\p'_{|[0,1]\times\{S\}\times [0,1]}\equiv S$, and  let us define $\s'$ by linear extension in $y\in[\min\p,S]$ elsewhere. This completes the proof.
\qed\end{pf}
\medskip

\begin{rem}{\em 
It is worth noting that in our construction, starting from a $2$-manifold,  the requirement  $\p'$ and $\s'$ of class $C^{0}$  cannot be improved to $C^{2}$.
Indeed, it is possible to construct examples of reduced size functions $\ell^*_{(\M,\p)}$ such that each point of the set
$$X:=\{(x,x)\in \Delta: 0\le x\le 1\}$$
is the limit point for a sequence of  cornerpoints of $\ell^*_{(\M,\p)}$.
If it were possible to construct $\p'$ of class $C^{2}$ such that $\ell^*_{(\M,\p)}=\ell^*_{(\M',\p')}$, the coordinates of each point in $X$ would be the limit point for a sequence of critical values of $\p'$ (cf. Cor.~2.3 in \cite{frosini96}). Since  the set of critical values is closed, it would follow that any point in $[0,1]$ is a critical value for $\p'$. This would contradict the Morse-Sard Theorem stating that, for a $C^r$ map $f$ from an $m$-manifold to an $n$-manifold, if $r>\max\{0,m-n\}$ then the set of critical values of $f$ has measure zero in the co-domain (cf. \cite{hirsch}, p.~69). We do not know whether it is possible to prove  Lemma~\ref{cubi} with $\p'$ and $\s'$ of class $C^{1}$ on a $2$-manifold.
Similarly, the Morse-Sard Theorem implies that it is not  possible to construct size pairs satisfying the properties of Lemma~\ref{cubi}, with $\p'$ and $\s'$ of class $C^{1}$ on a $1$-manifold, but we cannot exclude that this could be done by means of $C^{0}$ functions.}
\end{rem}

\section{Comparison with earlier results}

This section aims to show  that $d_{match}$ is the most suitable metric to compare reduced size functions,  mainly for two reasons. In the first place, we prove that one cannot find a distance between reduced size functions giving a better lower bound for the natural pseudo-distance than the matching distance. In the second place, we show that the lower bound provided by the matching distance improves an earlier estimate also based on size functions.
The main tool to obtain these results is Lemma~\ref{cubi}.

\begin{thm}\label{ermejo}
Let $\delta$ be a distance between reduced size functions, such that
$$\delta(\ell^*_{(\M,\p)},\ell^*_{(\N,\s)})\le \inf_{h\in H(\M,\N)}\max_{P\in\M}|\p(P)-\s(h(P))|,$$
for any two size pairs $(\M,\p)$ and $(\N,\s)$ with $\M$ and $\N$ homeomorphic.
Then,
$$\delta(\ell^*_{(\M,\p)},\ell^*_{(\N,\s)})\le d_{match}(\ell^*_{(\M,\p)},\ell^*_{(\N,\s)}).$$
\end{thm}

\begin{pf}
We argue by contradiction. Let us assume that there exist two size pairs $(\M,\p)$ and $(\N,\s)$, with $\M$ and $\N$ homeomorphic, such that
$$d_{match}(\ell^*_{(\M,\p)},\ell^*_{(\N,\s)})<\delta(\ell^*_{(\M,\p)},\ell^*_{(\N,\s)}).$$
By Lemma~\ref{cubi}, there exist $(\M',\p')$ and $(\M',\s')$ such that
$\ell^*_{(\M,\p)}=\ell^*_{(\M',\p')}$, $\ell^*_{(\N,\s)}=\ell^*_{(\M',\s')}$, and
$$d_{match}(\ell^*_{(\M',\p')},\ell^*_{(\M',\s')})=
\inf_{h\in  H(\M',\M')}\max_{P\in\M'}|\p'(P)-\s'(h(P))|.$$
Of course, $\delta(\ell^*_{(\M,\p)},\ell^*_{(\N,\s)})=\delta(\ell^*_{(\M',\p')},\ell^*_{(\M',\s')})$.
Hence,
\begin{eqnarray*}
\lefteqn{
\inf_{h\in  H(\M',\M')}\max_{P\in\M'}|\p'(P)-\s'(h(P))|= d_{match}(\ell^*_{(\M',\p')},\ell^*_{(\M',\s')})=}\\
& & = d_{match}(\ell^*_{(\M,\p)},\ell^*_{(\M,\s)})< \delta(\ell^*_{(\M,\p)},\ell^*_{(\M,\s)})=\\
& & = \delta(\ell^*_{(\M',\p')},\ell^*_{(\M',\s')})\le \inf_{h\in  H(\M',\M')}\max_{P\in\M'}|\p'(P)-\s'(h(P))|,
\end{eqnarray*}
giving a contradiction.
\qed\end{pf}

Analogously, we show that
the inequality of Th.~\ref{nsd}
$$d_{match} (\rsfp,\rsfn)\le \inf_{h\in H(\M,\N)}\max_{P\in\M}|\p(P)-\s(h(P))|$$
is a better bound than the one given in~\cite{Donatini}, that we restate here for reduced size functions:

\begin{thm}\label{fd}
 If there exist $(x,y)$ and $(\xi,\eta)$ in $\Delta^+$ such that $\rsfp(x,y)>\rsfn(\xi,\eta)$, then
$$\inf_{h\in H(\M,\N)}\max_{P\in\M}|\p(P)-\s(h(P))|\ge\min\{\xi-x,y-\eta\}.$$
\end{thm}

Indeed, we have the following result.

\begin{thm}\label{confronto}
Assume that
$$A:=\left\{\left((x,y),(\xi,\eta)\right)\in\Delta^+\times\Delta^+: \xi\ge x, \eta\le y, \rsfp(x,y)>\rsfn(\xi,\eta) \right\}$$
is non-empty, and let
$$s:=\sup_{\left((x,y),(\xi,\eta)\right)\in A}\{\min\{\xi-x,y-\eta\}\}$$
(in other words, $s$ is the best non-negative lower bound we can get for the natural pseudo-distance $\inf_h\max_{P\in\M}|\p(P)-\s(h(P))|$ by applying Th.~\ref{fd}).
Then $$d_{match}(\rsfp,\rsfn)\ge s.$$
\end{thm}

\begin{pf}
 We argue, by contradiction, assuming $d_{match}(\rsfp,\rsfn)<s$.
Since $s$ is a $\sup$, there is a pair
$\left((\bar x,\bar y),(\bar \xi,\bar \eta)\right)\in A$  satisfying
$$d_{match}(\rsfp,\rsfn)< \min\{\bar \xi-\bar x,\bar y-\bar \eta\}=:s'$$
with $s'\le s$.

By Lemma~\ref{cubi} we can construct two size pairs $(\M',\p')$ and
$(\M',\s')$ such that $\ell^*_{(\M,\p)}=\ell^*_{(\M',\p')}$, $\ell^*_{(\N,\s)}=\ell^*_{(\M',\s')}$, and
$$d_{match}(\ell^*_{(\M',\p')},\ell^*_{(\M',\s')})=
\inf_{h\in H(\M',\M')}\max_{P\in\M'}|\p'(P)-\s'(h(P))|.$$
Clearly, the set $A$ coincides with the set
$$B:=\left\{\left((x,y),(\xi,\eta)\right)\in\Delta^+\times\Delta^+: \xi\ge x, \eta\le y, \ell^*_{(\M',\p')}(x,y)>
\ell^*_{(\M',\s')}(\xi,\eta) \right\}.$$
Then $\left((\bar x,\bar y),(\bar \xi,\bar \eta)\right)\in B$, implying
$s'\le \inf_h\max_{P\in\M'}|\p'(P)-\s'(h(P))|$, because of Th.~\ref{fd}.
Finally, observing that
\begin{eqnarray*}
\lefteqn{s' \le \inf_{h\in H(\M',\M')}\max_{P\in\M'}|\p'(P)-\s'(h(P))|=}\\
& & =  d_{match}(\ell^*_{(\M',\p')},\ell^*_{(\M',\s')})= d_{match}(\rsfp,\rsfn)<s',
\end{eqnarray*}
we obtain  a contradiction.
\qed\end{pf}

\section{Conclusions}
The main contribution of this paper is the proof that an appropriate distance between reduced size functions, based
on optimal matching, provides the best, stable and easily computable lower
bound for the natural pseudo-distance between size pairs. Hence,
 the problem of estimating the dissimilarity between size pairs by the natural pseudo-distance can be
dealt with by matching the points of the representative sequences of reduced size functions. The estimate in
Th.~\ref{nsd} improves an earlier one  given in \cite{Donatini}, also based on  size functions but without the use
of cornerpoints.

The stability of the matching distance between reduced size functions with respect to continuous functions is important by itself.
Indeed, it allows us to use reduced size functions as shape descriptors with the confidence that they are robust against
perturbations on the data, often arising in real applications due to noise or errors.

A crucial result in our paper is the proof that it is always possible to construct two suitable measuring functions on a topological $2$-sphere
with given reduced size functions and a pseudo-distance
equaling their matching distance (Lemma \ref{cubi}). This result has allowed us to prove that the matching distance is the best tool to compare reduced size functions. Indeed, after Th. \ref{ermejo}, we know that it makes no sense to look for different metrics on size functions in order to improve the lower bound for the natural pseudo-distance furnished by the matching distance. However, it would be interesting to study whether  the inequality of Theorem~10 in \cite{FroMu99}, giving a lower bound for the natural pseudo-distance via the size homotopy groups, can be improved  using appropriate analogs of the concepts of cornerpoint and  matching distance, in the same way that the estimate in
Th.~\ref{nsd} improves the result  given in
\cite{Donatini}. 

\bibliographystyle{siam}
\bibliography{sf_biblio070702}

\end{document}